\documentclass[useAMS,usenatbib,fleqn]{mn2e}
\pdfoutput=1
\usepackage[usenames,dvipsnames]{color}
\usepackage{times}
\usepackage{mathptmx}
\usepackage{graphicx}
\usepackage{gensymb}
\usepackage[table]{xcolor}
\usepackage{etoolbox}
\usepackage{xspace}
\usepackage{paralist}
\usepackage[usenames,dvipsnames]{color}
\usepackage{amsmath, bm}

\title[Stellar and total mass in the Galactic bulge region]{Made-to-Measure models of the Galactic Box/Peanut bulge: \\ stellar and total mass in the bulge region}

\author[M. Portail et al.]
  {M.~Portail,$^1$\thanks{E-mail:portail@mpe.mpg.de}
  C.~Wegg,$^1$ O.~Gerhard,$^1$ and I.~Martinez-Valpuesta$^{2,3}$\\
  $^1$ Max-Planck-Instit\"{u}t f\"{u}r Extraterrestrische Physik, Gie\ss enbachstra\ss e, D-85741 Garching, Germany\\
  $^2$ Instituto de Astrof\'{i}sica de Canarias, E-38205 La Laguna, Tenerife, Spain\\
  $^3$ Universidad de La Laguna, Dept. Astrof\'{i}sica, E-38206 La Laguna, Tenerife, Spain}

\begin{document}
\renewcommand{\arraystretch}{1.3}

\renewcommand{\vec}[1]{\boldsymbol{#1}}
\newcommand{\pc}{\,\rm{pc}\xspace}
\newcommand{\kpc}{\,\rm{kpc}\xspace}
\newcommand{\kms}{\,\rm{km\,s^{-1}}\xspace}
\newcommand{\kmskpc}{\,\rm{km\,s^{-1}\,kpc^{-1}}\xspace}
\mathchardef\mhyphen="2D

\newcommand{\nmagic}{\textsc{nmagic}\xspace}
\newcommand{\brava}{\textsc{brava}\xspace}
\newcommand{\argos}{\textsc{argos}\xspace}
\newcommand{\twomass}{\textsc{2mass}\xspace}
\newcommand{\ukidss}{\textsc{ukidss}\xspace}
\newcommand{\glimpse}{\textsc{glimpse}\xspace}
\newcommand{\ogle}{\textsc{ogle}\xspace}
\newcommand{\vvv}{\textsc{vvv}\xspace}

\newcommand{\iu}{\,\rm{iu}\xspace}
\newcommand{\masyr}{\,\rm{mas\,yr^{-1}}\xspace}
\newcommand{\iteration}{\,\rm{it}\xspace}
\newcommand{\SM}{10^{10}\rm{M}_{\odot}\xspace}
\newcommand{\Msun}{\,\rm{M}_{\odot}\xspace}
\newcommand{\bb}{\rm{b\mhyphen b}}
\newcommand{\R}{\mathcal{R}}
\newcommand{\Ms}{M_{\rm{s}}}
\newcommand{\MDM}{M_{\rm{DM}}}
\newcommand{\Mtot}{M_{\rm{tot}}}
\newcommand{\Gyr}{\,\rm{Gyr}}
\newcommand{\F}{\mathcal{F}\xspace}
\newcommand{\vel}{\textrm{v}}

\newcommand{\reftable}{\rm{Table}\xspace}
\newcommand{\reffigure}{\rm{Fig.}\xspace}
\newcommand{\refequation}{\rm{Eq.}\xspace}
\newcommand{\refequations}{\rm{Eqs.}\xspace}
\newcommand{\refsection}{\rm{Section}\xspace}

\definecolor{light-gray}{gray}{0.80}

\date{Accepted 2015 January 9.  Received 2014 December 12; in original form 2014 August 23}

\pagerange{\pageref{firstpage}--\pageref{lastpage}} \pubyear{2015}

\maketitle


\label{firstpage}

\begin{abstract}

\noindent We construct dynamical models of the Milky Way's Box/Peanut (B/P) bulge, using the recently measured 3D density of Red Clump Giants (RCGs) as well as kinematic data from the \brava survey. We match these data using the \nmagic Made-to-Measure method, starting with N-body models for barred discs in different dark matter haloes. We determine the total mass in the bulge volume of the RCGs measurement ($\pm 2.2 \times \pm 1.4 \times \pm 1.2\kpc$) with unprecedented accuracy and robustness to be $1.84 \pm 0.07 \times \SM$. The stellar mass in this volume varies between $1.25-1.6 \times \SM$, depending on the amount of dark matter in the bulge. We evaluate the mass-to-light and mass-to-clump ratios in the bulge and compare them to theoretical predictions from population synthesis models. We find a mass-to-light ratio in the K-band in the range $0.8-1.1$. The models are consistent with a Kroupa or Chabrier IMF, but a Salpeter IMF is ruled out for stellar ages of $10 \Gyr$. To match predictions from the Zoccali IMF derived from the bulge stellar luminosity function requires $\sim 40\%$ or $\sim 0.7 \times \SM$ dark matter in the bulge region. The \brava data together with the RCGs 3D density imply a low pattern speed for the Galactic B/P bulge of $\Omega_{\rm p} = 25-30 \kmskpc$. This would place the Galaxy among the slow rotators ($\R \geq 1.5$). Finally, we show that the Milky Way's B/P bulge has an off-centred X structure, and that the stellar mass involved in the peanut shape accounts for at least $20\%$ of the stellar mass of the bulge, significantly larger than previously thought.

\end{abstract}


\section{Introduction}
\label{section:introduction}

Observations of external disc galaxies have shown that about half of all disc galaxies have strong bars \citep{Eskridge2000}. The Milky Way Galaxy (MW) has been considered for many years as one of them. The Galactic bar/bulge causes the non-circular motions in the gas flow seen in H{\scriptsize I} and CO \citep{Peters1975,Binney1991,Englmaier1999,Fux1999} and is the origin of the asymmetries seen in the near-infrared light distribution \citep{Blitz1991,Weiland1994,Binney1997} and star counts \citep{Nakada1991,Stanek1997,Lopez-Corredoira2005}.  The Galactic Bulge (GB) is regarded as the three-dimensional part of the bar seen nearly end-on \citep{Shen2010, Martinez-Valpuesta2011} with a semi-major axis of about $2\kpc$ \citep{Gerhard2002a, Wegg2013}, as is also indicated by the near-cylindrical rotation of the bulge stars \citep{Beaulieu2000,Kunder2012,Ness2013a}.

In the last decade, stellar surveys of the GB such as \twomass \citep{Skrutskie2006}, \vvv \citep{Saito2012}, \ogle \citep{Sumi2004}, \brava \citep{Rich2007} and \argos \citep{Freeman2013} have released unprecedented data sets which allow us to study the GB star-by-star. The triaxial bulge of the MW is now believed to be a so called Box/Peanut bulge (B/P bulge) or X-shaped bulge as indicated by its bimodal distribution of Red Clump Giants (RCGs). This was first reported from the \twomass catalogue by \citet{McWilliam2010} and from the \textsc{ogle-iii} survey by \citet{Nataf2010}. \citet{Ness2012} showed that this split red clump is seen for stars with metallicity $[\rm{Fe/H}] > -0.5$. The peanut shape was mapped last year in three dimensions by \citet{Wegg2013} using public data from the \vvv survey.

Star counts and infrared observations have revealed a long and flat bar component, located mostly in the Galactic plane and extending up to $l\sim27\degree$ \citep{Hammersley2000,Benjamin2005, Cabrera-Lavers2007}. Curiously, the angle of the bar relative to the line-of-sight to the Galactic Centre (GC) was inferred to be $\phi = 45\degree$ in these studies, while the angle of the barred B/P bulge is accurately measured from the RCGs as $\phi = 27\degree \pm 2\degree$ \citep{Wegg2013}. Study of N-body models suggested that the long bar and the B/P bulge data could be explained by a unique B/P bulge and bar structure formed by the buckling instability, if its two-dimensional outer bar component has developed leading ends through interaction with adjacent spiral arm heads \citep{Martinez-Valpuesta2011}. 

The buckling instability is a well studied process in N-body simulations \citep{Combes1981, Raha1991, Martinez-Valpuesta2004}. Cold stellar discs embedded in live dark haloes naturally tend to form a bar which experiences long term secular evolution because of angular momentum transfer from the disc to the halo \citep{Debattista2000,Athanassoula2003}. During this secular evolution the bar can go through one or more buckling events \citep{Raha1991, Martinez-Valpuesta2006}, making it vertically thick and creating the so called Box/Peanut shape, or X-shape in unsharp-masked images.

In this context, the goal of this paper is to combine N-body simulations of evolved stellar discs with recent MW data to create dynamical models of the Galactic bulge and study its total mass, mass-to-light ratio, stellar mass and X-shape structure. To do so we use the 3D density of RCGs from \citet[][hereafter also WG13]{Wegg2013} as well as kinematic data from the \brava survey \citep{Howard2008, Kunder2012} to create particle models using the Made-to-Measure method (M2M). In M2M modelling, the weights of the particles in an N-body system are continuously updated until the observables from the model match a set of target data constraints. The M2M method was introduced by \citet{Syer1996} and recast by \citet{DeLorenzi2007} such that observational errors can be taken into account. We use the \nmagic code of \citet{DeLorenzi2007} adapted to the Milky Way problem by \citet{Martinez-Valpuesta2012}. \nmagic has been used mostly in extragalactic studies \citep{DeLorenzi2009,Das2011,Morganti2013}. M2M methods were previously used in the Milky Way context by \citet{Bissantz2004}, \citet{Long2013}, also using \brava data, and \citet{Hunt2014}.

This paper is organized as follows. In \refsection \ref{section:particleModels}, we present the set of N-body simulations of barred discs that we use as starting point for our M2M modelling. In \refsection \ref{section:M2M}, we recount the M2M method and the data sets we use to model the GB. The modelling is detailed in \refsection \ref{section:DynamicalModels} where we present our best dynamical models and discuss the issue of the pattern speed of the bar. In \refsection \ref{section:mass}, we show that we can recover the total mass in the bulge region with great accuracy, thereby relating the stellar mass to the dark matter mass in the bulge.  In \refsection \ref{section:MtoLAndRCGDensity}, we compute the stellar mass-to-light and mass-to-clump ratios of our models and compare them to theoretical predictions from population synthesis models. Section \ref{section:Xshape} quantifies the importance of the X-shape structure in the bulge which accounts for more than 20\% of the stellar mass of the bulge. Finally, we discuss our results in \refsection \ref{section:Discussion} and summarize in \refsection \ref{section:Conclusion}.

\section{Particle models of barred discs}
\label{section:particleModels}

\subsection{N-body models in different dark matter haloes}
\label{section:setuphaloes}

Our M2M modelling of the GB relies on reasonable initial particle models of barred discs. These initial models were created by evolving near-equilibrium stellar discs embedded in live dark matter haloes. Near-equilibrium models are constructed using the program \textsc{magalie} \citep{Boily2001} and evolved with the tree-code \textsc{gyrfalcon} \citep{Dehnen2000}, all distributed with the publicly available \textsc{nemo} toolbox \citep{Teuben1995}. During the evolution, the disc naturally forms a bar which rapidly buckles out of the Galactic plane and creates a B/P bulge \citep{Combes1981, Raha1991}.

As we want to address the question of the total mass of the GB including the amount of dark matter in the bulge, we generated a set of five disc+halo N-body models using the same disc component and varying the halo properties. The disc is exponential with scale length of $1$ internal units (iu), scale height of $0.14 \iu$, unit total disc mass, and $Q$ parameters of $1.4$ at $R = 3.07\iu$. Halos have a Hernquist density profile with flattening of $0.8$ and a sharp cutoff at $20 \iu$. All models contain two million particles, one million each for disc and halo. With these settings fixed only two free parameters remain: the halo mass inside the cutoff, $M_h$, and the halo break radius of the Hernquist profile, $a_h$. They were fixed considering the total rotation curve of a model. Following the language of \citet{Sackett1997}, we call the \emph{degree of maximality} of a disc the proportion of the disc contribution to the total velocity curve, at the radius where the disc velocity curve is maximal. Assuming a flat rotation curve at large radii, we build a one parameter family of models parametrized by the degree of maximality of the disc. We make five models of this family, called M80, M82.5, M85, M87.5 and M90 which have different degree of maximality ranging from $80 \%$ to $90\%$. As we show later this allows us to build models of barred galaxy which span all kind of rotators, from slow to fast. The halo parameters used to construct these models are summarized in \reftable \ref{table:modelsParameters}. 

\begin{figure}
  \centering
  \includegraphics{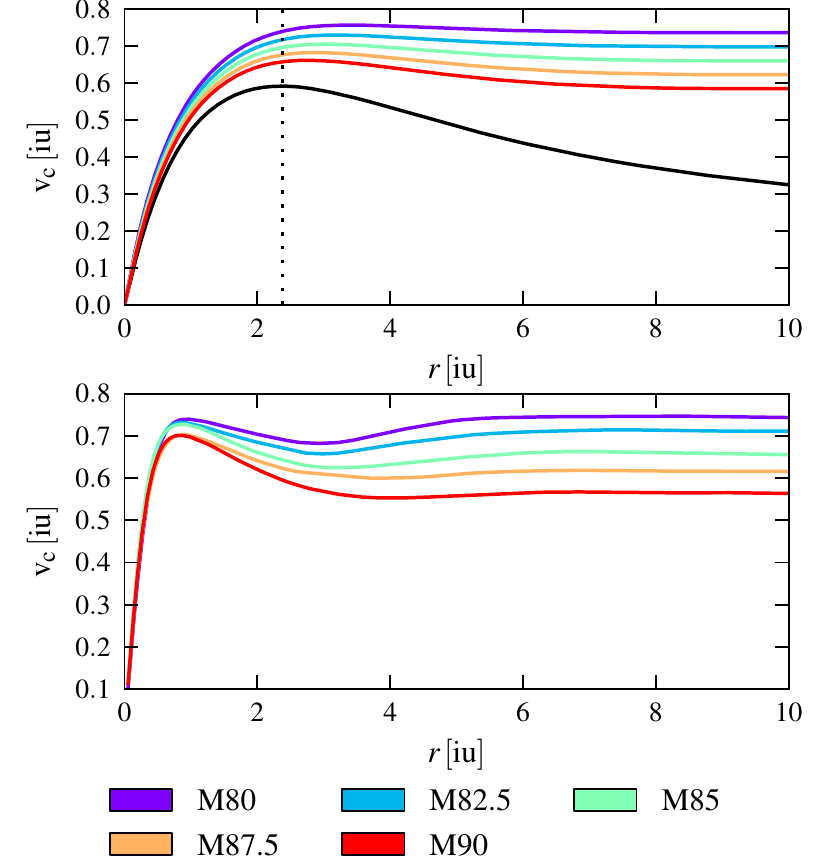}\\
  \caption{\emph{Top :} Circular velocity curves of our five disc-halo models before evolution, in model units. The black line shows the circular velocity curve of the disc only, kept the same for all models. The dashed vertical line indicates the radius where the maximality of the disc is computed. \emph{Bottom :} Azimuthally averaged circular velocity curves after the bar and B/P bulge formation, also in model units.}
  \label{fig:circularVelocity}
\end{figure}

For each model, we select a snapshot of a late evolutionary stage, where the bar is fully grown. The circular velocity curves obtained from the azimuthally averaged potential of this set of models are plotted in \reffigure \ref{fig:circularVelocity}, before any evolution (top) and, for the selected snapshot after bar and B/P bulge formation (bottom). The different colours refer to different models as stated in the legend. In the upper plot, the solid black line is the circular velocity curve of the initial disc component (common for all models) and the vertical dashed line indicates the radius at which the degree of maximality is determined. As expected the bar formation moves material inward which increases the circular velocity in the inner region of the disc.

\subsection{Geometry and scaling}
\label{section:geometryAndScaling}

 According to the latest studies \citep{Chatzopoulos2014, Reid2014}, we assume a distance to the Galactic Centre (GC) of $R_{0} = 8.3\kpc$. The bar is placed at an angle of $\alpha = 27\degree$ from the Sun-GC line (WG13). All our models are scaled independently based on the length of their long bar. Several studies based on different data sets indicate that the long bar of the MW ends at about $l = 27 \degree$. \citet{Hammersley2000} studied star counts of $K$ giants in several in-plane fields and places the end of the long bar at $l \sim 27\degree$ while \citet{Cabrera-Lavers2007, Cabrera-Lavers2008} found respectively values of $27\degree$ and $28\degree$, based on RCGs counts from the \twomass catalogue and from \ukidss data. \citet{Benjamin2005} also found indication of a long bar for $l < 30\degree$ using star counts from the \glimpse catalogue. All these different estimates agree well with each other and give therefore a consistent length scale. With $R_{0} = 8.3\kpc$, a bar angle of $27\degree$ and the end of the bar at $l=27\degree$, the half length of the bar is $R_{\rm{bar}} = 4.66\kpc$. The different parameters quoted above constitute our fiducial set of parameters. We checked that our main results are not significantly affected by assuming a bar angle of $32\degree$ instead of $27\degree$ or a shorter bar half-length of $R_{\rm{bar}} = 3.8\kpc$ (see {\refsection \ref{section:Systematics}}), or by using $R_{0} = 8\kpc$ instead of $8.3\kpc$.
 
 We used ellipse fitting of the face-on projection of our models to compute their bar lengths, and scaled them to $4.66\kpc$. The velocity scaling is kept free and will be determined dynamically from the data during the modelling process, as explained in \refsection \ref{section:dynamicalScaling}.
 
 In order to avoid referring to scale dependent quantities we will refer to the rotational velocity of the bar using the dimensionless number $\R = R_\textrm{cr}/R_\textrm{bar}$, the ratio between the corotation radius and the half-length of the bar. Bars with $\R \geq 1.4$ are called slow rotators, while those with $\R \leq 1.4$ are fast rotators \citep{Debattista2000}. Because the bar cannot extend beyond corotation \citep{Contopoulos1980}, $\R$ has to be larger than one. Our models span $\R$ values quite uniformly from $1.8$ to nearly $1.1$, which corresponds to the full range of reasonable values for barred galaxies \citep{Elmegreen1996a}. Statistics for external galaxies from \citet{Rautiainen2008} show that nearly all galaxies of Hubble types S0, SBa or SBb are consistent with being fast rotators while for SBc galaxies (like the MW), bars can be either fast or slow. The sampling of $\R$ values we obtain is a consequence of the different haloes we used. The halo absorbs angular momentum during the bar formation and therefore the more maximal the initial disc, the less halo material there is to absorb angular momentum and so the faster the bar. One should keep in mind that for our set of models, the halo mass in the inner parts and the $\R$ value are not independent parameters. The different model and bar parameters are given in \reftable \ref{table:modelsParameters}.

\begin{figure}
  \includegraphics{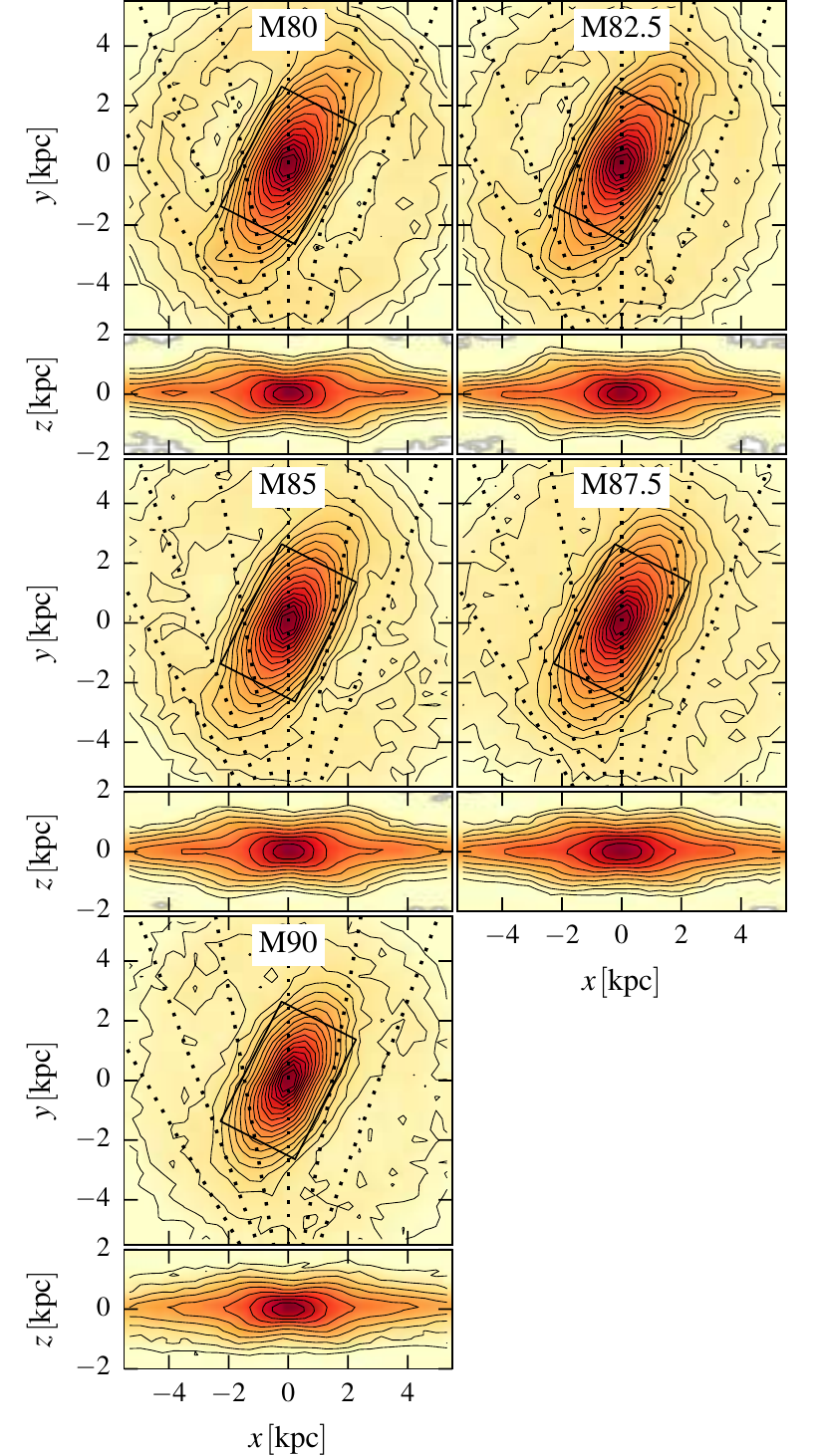} \\
  \caption{Face-on and side-on views of our five initial models of barred discs with B/P bulges. The Sun is located $8.3\kpc$ from the Galactic center and the bar is at an angle of $27\degree$ from the Sun-GC line. The dotted lines on the face-on view indicate sight lines spaced every $10\degree$ in Galactic longitude. The bold rectangle indicates the boundary of the box where the density is given by \citet{Wegg2013}, see \refsection \ref{section:M2MDensityObservable}.}
  \label{fig:modelsPlots}
\end{figure}

Throughout this paper the $(x,y,z)$ frame refers to the Galactocentric inertial frame where $z$ is the vertical axis pointing to the Galactic North and $y$ the Sun-GC axis. The bar rotates at pattern speed $\Omega_{\rm p}$ in this inertial frame and we shall refer to the rotating frame of the bar as $(x',y',z')$, with $x',y'$ and $z'$ respectively the major, intermediate and vertical axis of the bar. \reffigure \ref{fig:modelsPlots} shows the face-on and side-on projections of our models with these geometry and scaling.

\begin{table}
\caption{Main parameters of our five disc-bar-halo models. The two first rows show the initial halo parameters: the halo mass inside cutoff $M_h$ in units of the disc mass $M_d$ and the Hernquist break radius $a_h$. The next rows give parameters after bar and bulge formation and scaling to physical units (see \refsection \ref{section:geometryAndScaling}): the corotation radius $R_\textrm{cr}$ of the bar, the ratio of the corotation radius over bar half-length $\R$ and the dark matter fraction in the bulge $\MDM/\Mtot$.}
\label{table:modelsParameters}
  \centering
  \begin{tabular}{l|ccccc}
		  & M80 & M82.5 & M85 & M87.5 &M90 \\
      \hline\hline
      $M_h / M_d$ & 9.42 & 8.51 & 7.68 & 6.94 & 6.27 \\
      $a_h [\iu]$ & 17.62 & 19.44 & 22.05 & 26.16 & 33.68 \\
      $R_\textrm{cr}\,[\rm{kpc}]$  & 8.4 & 7.6 & 6.7 & 5.6 & 5.0 \\
      $\R = R_\textrm{cr}/R_\textrm{bar}$  & 1.80 & 1.64 & 1.44 & 1.21 & 1.08 \\
      $\MDM/\Mtot$ & 44.2\% &  40.8\% & 34.5\%  & 28.6\%  & 25.2\% \\
  \end{tabular} 
\end{table}

\section{Made-to-measure with MW observables}
\label{section:M2M}
As an alternative to distribution function-based methods \citep{Dejonghe1984, Binney2010}, moment-based methods \citep{Binney1990, Cappellari2009} or orbit-based methods \citep{Schwarzschild1979, Thomas2009}, \citet{Syer1996} proposed a particle-based algorithm to study stellar dynamical equilibria, known as the Made-to-Measure (M2M) method. This algorithm consists of adapting the particles weights of an initial particle model of the system of interest such as to reproduce a given set of observables. \citet[hereafter DL07]{DeLorenzi2007} improved the original method by \citet{Syer1996} to take  into account observational errors, and implemented it as the \nmagic code. \nmagic has been used in numerous studies, mostly in the context of elliptical galaxies \citep[e.g.][]{DeLorenzi2009,Das2011,Morganti2013} and has been adapted to barred galaxies by \citet{Martinez-Valpuesta2012}. In the context of the Milky Way, the M2M method has previously been used by \citet{Bissantz2004}, \citet{Long2013} and \citet{Hunt2014}.

\subsection{Theory of the M2M method}
\label{section:M2MTheory}
Let us consider a system characterized by its DF $f(\vec{z})$ defined on the phase space of the system. Any observable $y_j$ of this system will be written as
\begin{equation}
 y_j = \int K_j(\vec{z}) f(\vec{z}) \,d^6z, 
\end{equation}
where $K_j$ is the kernel of the observable and $\vec{z}$ the phase space vector. If we represent the system by a set of $N$ particles, with particle weights $w_i(t)$, the observable will be written as 
\begin{equation}
\label{equation:observable}
 y_j(t) = \sum_{i=1}^N K_{j}(\vec{z}_i(t)) w_i(t), 
\end{equation}
where $\vec{z}_i(t)$ is the phase-space coordinate of particle $i$ at time $t$. The $w_i$ are proportional to the physical weights of the particles but one can also see them as density-elements of the phase-space.

Let us now consider different data sets, indexed by the subscript $k$, that one wants to fit. The difference between the model and the observational target is quantified using the residuals $\Delta_j^k(t)$ defined as

\begin{equation}
\label{equation:delta}
\Delta_j^k(t) = \frac{y_j^k(t) - Y_j^k}{\sigma(Y_j^k)}, 
\end{equation}
where $Y_j^k$ denotes an observable of data set $k$, $\sigma(Y_j^k)$ its associated error, and $y_j^k(t)$ the corresponding model observable.

The M2M method will adapt the weights of the particles in order to maximize the profit function $F$ defined by

\begin{equation}
 F = \mu S - \frac{1}{2} \chi_{\rm{tot}}^2, 
 \label{equation:profit}
\end{equation}
where
\begin{equation}
 S = -\sum_i w_i \log\left(\frac{w_i}{\widehat{w_i}}\right)
 \label{equation:entropy}
\end{equation}
and 
\begin{equation}
 \chi^2_{\rm{tot}} = \sum_{k,j} \lambda_k(\Delta_j^k)^2~.
 \label{equation:chi2}
\end{equation}

\noindent$S$ corresponds to an entropy term used to regularize the particle weights in order to ensure that they do not deviate too much from a set of predefined priors $\widehat{w_i}$. The $\lambda_k$ are numerical weights of the different data sets, as formally introduced by \citet{Long2010}. The determination of these factors is discussed in \refsection \ref{section:M2Mlambda}.

The heart of the M2M method is the following weight evolution equation:
\begin{align}
  \label{equation:FOC1}
  \frac{dw_i}{dt} &= \varepsilon w_i(t) \left[ \frac{\partial F}{\partial w_i}\right]\\
 \label{equation:FOC2}
 &= \varepsilon w_i(t) \left[ \mu \frac{\partial S}{\partial w_i} - \sum_k \lambda_k \sum_{j}  \frac{K_j(\vec{z}_i(t))}{\sigma(Y_j^k)} \Delta_j^k(t) \right]
\end{align}
where the bracket term is the so-called force-of-change. Note that the passage from \refequation \ref{equation:FOC1} to  \refequation \ref{equation:FOC2} is made under the assumption that the kernels $K_j^k$ do not depend on the weights of the particles. We will make this assumption as it allowed \citet{Syer1996} and DL07 to prove the convergence of the particle weights for small linear deviations from the solution. No additional term is used in \refequation \ref{equation:FOC2}, in particular we do not re-normalize the weights of the particles. This is done on purpose to allow fitting different masses of the bulge as shown in \refsection \ref{section:DynamicalModels}.

In order to reduce the shot noise of the particle model we follow \citet{Syer1996} and DL07 and artificially increase the effective number of particles using temporal smoothing, replacing $y_j(t)$ in \refequation \ref{equation:delta} by
\begin{equation}
\tilde{y}_j(t) = \int y_j(t-\tau)e^{-\alpha \tau} \, d\tau~.
\end{equation}

\subsection{Density observables}
\label{section:M2MDensityObservable}
\subsubsection{3D density of Red Clump Giants}

Using the Red Clump Giants (RCGs) from the \vvv survey, WG13 measured the three-dimensional density distribution of the inner part of the Galactic bar/bulge. They evaluate line-of-sight density distributions of the RCGs by deconvolution of extinction and completeness corrected $K_s$ band magnitude distributions for different \vvv fields. To do so, they fit a background to each $K_s$ band magnitude distribution and identify the RCGs as the excess over this background. Assuming an 8-fold mirror symmetry, they constructed a 3D density map of the RCGs in the inner $\pm 2.2 \times \pm 1.4 \times \pm 1.2\kpc$  along the $(x',y',z')$ axes of the galactic bar/bulge, and checked that the data show only small deviations from 8-fold symmetry (their \reffigure 15). Their work relies on several assumptions that they carefully investigated for systematic variations of their fiducial parameters. They finally provide us with one fiducial and five variant density maps of RCGs in the bulge. We use the range of these variant maps as a systematic error on the density measurements. The typical magnitude of these errors is about $10\%$.

We assume here that RCGs are good tracers of the stellar mass in the bulge, i.e. that the ratio of number of RCGs per unit of stellar mass stays constant in the bulge. This assumption is supported by the theoretical work of \citet{Salaris2002} where they showed that for an old star population of about $10\, \rm{Gyr}$ one would expect the number of RCGs to vary by less that $10\%$ for metallicity in the range $-1.5 \leq [M/H]\leq 0.2$. As the bulge appears uniformly old with no significant metallicity component extending beyond the range $-1.5 \leq [M/H]\leq 0.2$ \citep{Zoccali2003}, we can consider RCGs as good tracers of the stellar mass and therefore use the map of WG13 as a constraint on the shape of the stellar mass density in our models. Note that the map gives only the shape of the density and not its absolute value because the number of RCGs per unit of stellar mass is a priori unknown. 

Unfortunately high extinction and crowding prevented WG13 to reliably measure the RCGs density within $150\pc$ of the Galactic plane, and also cause some uncertainty immediately above this $150\pc$ strip. Using directly the original map with this missing strip would be inappropriate since we want to model different bulge masses by changing the scaling of the density constraint. An incomplete map would let the midplane free and lead to unrealistic models, with for example a light in-plane component and a massive out-of-plane component. Hence we model the density in the missing $\pm 150\pc$ strip by extrapolation of each vertical density profile of the original 3D map.

We found that the vertical density profiles in both the data of WG13 and the initial particle models are well represented by a $\rm{sech}^2$ function. Our fiducial extrapolation is thus based on the best $\rm{sech}^2$ fit of each vertical profile. We account for the uncertainty due to the choice of the extrapolation law by considering a different one in \refsection \ref{section:Systematics}, showing that the total bulge mass is insensitive to this choice.

\begin{figure}
  \includegraphics{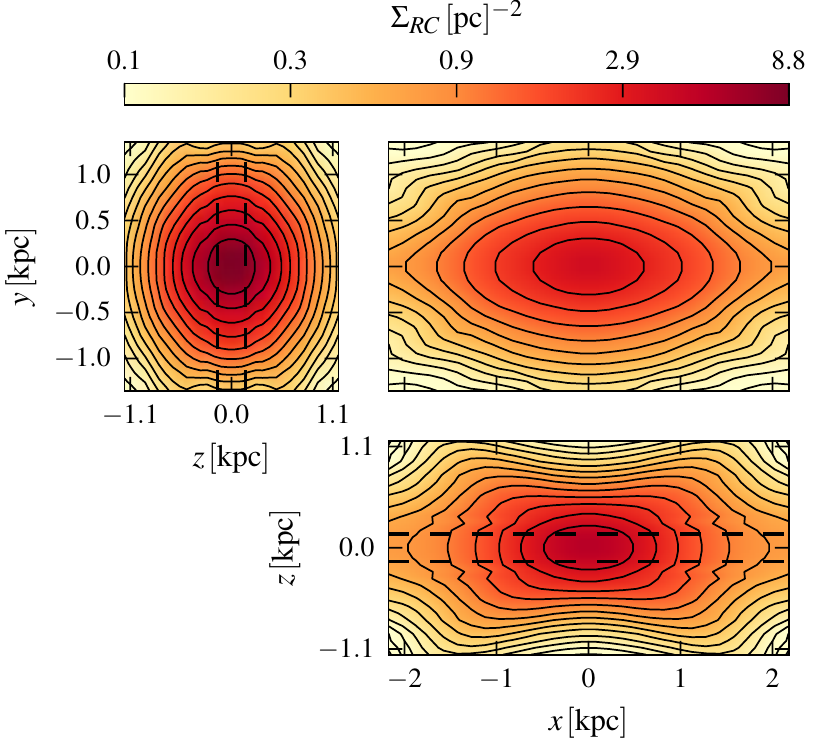}
  \caption{End-on (top left), face-on (top right) and side-on (bottom right) projection of the extrapolated 3D RCGs density map originally from \citet{Wegg2013}. The dashed lines in the end-on and side-on view show the $\pm 150 \pc$ region where the density map was extrapolated.}
    \label{fig:RCGDensity}
\end{figure}

The fiducial extrapolated density map of RCGs in the $\pm 2.2 \times \pm 1.4 \times \pm 1.2\kpc$ box is plotted in projection in \reffigure \ref{fig:RCGDensity}. The side-on view shows a very strong peanut-shape.

\subsubsection{Implementation in \nmagic}

The 3D density map was evaluated on a regular Cartesian grid of $(30,28,32)$ cells along the $(x',y',z')$ axis, covering a box of the inner $\pm 2.2 \times \pm 1.4 \times \pm 1.2 \kpc$, as in WG13. We will refer to this region as the ``bulge-in-box'' or the abbreviation $\bb$. From WG13 we know the total number of RCGs in the $\bb$, but the corresponding stellar mass must still be determined from dynamical modelling. We parametrize the total stellar mass of the $\bb$ using a dimensionless factor, $\F$,  defined as the ratio of the target stellar mass of the $\bb$ divided by its value in the initial model. The target density observables $ Y_j^{\rm{d}}$, corresponding to the target stellar mass in cells $j$, in internal units, are then given by
\begin{equation}
\label{equation:densityKernel}
 Y_j^{\rm{d}} = \F \, \left(\sum_{i \in \bb} w_i(t=0) \right ) \, \frac{n_\textrm{RCGs} (j)\, \Delta^3x}{\int_{\bb} n_\textrm{RCGs}\,d^3x}
\end{equation}
where the sum is over all initial particle weights in the $\bb$, $n_\textrm{RCGs}$ in the number density of RCGs, and $\Delta ^3x$ the volume of the cells. For a given scaling from model internal units to physical units, a change in the value of $\F$ is equivalent to a change in the target stellar mass of the $\bb$. For a given $\F$, \nmagic takes care of increasing or decreasing the weights of the particles to reach the target observables $Y_j^{\rm{d}}$, thereby modelling different-mass bulges.

The model density observables in the cells of the $\bb$ are then computed from {\refequation \ref{equation:observable}} using the following kernel
\begin{equation}
 K_j^{\rm{d}}  (\vec{z}_i) = 
 \begin{cases}
   1 & \text{if } i \in \text{cell }j,  \\
   0 & \text{otherwise}.
  \end{cases}
\end{equation}

\subsection{Kinematic observables}
\label{section:M2MkinematicObservable}
\subsubsection{\brava data}

We kinematically constrain our models using data from the Bulge RAdial Velocity Assay (\brava) \citep{Rich2007, Howard2008, Kunder2012}. The \brava survey is a large spectroscopic survey of M giant stars selected from the 2MASS catalogue. According to \citet{Howard2008}, the light of these M giants traces the $2 \micron$ light of the GB and therefore M giants are a good proxy to study the overall kinematics of the GB. The \brava survey provides us with galactocentric rest-frame mean radial velocity and velocity dispersion in more than $80$ fields through the bulge, mostly between $l = -10\degree$ to $l=10\degree$. On starting this project the \argos data \citep{Ness2013a} were not yet available so we restricted ourselves to the \brava data. The \argos data have a more complicated selection function and will be included in a later paper. 

Before constraining a model it is important to study in detail how the \brava stars were selected in order to reproduce any selection bias. The \brava stars were selected from their location on a $K$ versus $J-K$ colour-magnitude diagram aiming to select only bulge members with no metallicity bias. \citet{Howard2008} deployed a lot of effort to ensure no metallicity bias but the bulge membership criteria based on magnitude cuts adjusted by eye are more questionable. We used the \textsc{galaxia} model \citep{Sharma2011} to evaluate the possible foreground contamination using the same selection criteria. We found that contamination was negligible toward the centre but appears to rise with increasing Galactic longitude, reaching $20\%$ of the sample at $l\sim15\degree$. We found no significant variation with latitude in the latitude range of the \brava fields. We decided not to simulate this contamination, given the fact that the foreground discs in our models are not made to match the disc of the Milky Way. Instead, we simply exclude from our analysis all the fields outside the inner $\pm 10\degree$ in longitude, where contamination is probably significant. 
The magnitude cuts also introduce a slight bias towards the near side of the bulge. Faint M giants are more numerous than bright ones so have a larger probability to be part of the sample on the near side than on the far side of the GB. We model this effect below.

\subsubsection{Implementation on \nmagic}

As stated in \refsection \ref{section:M2MTheory}, the linear convergence of the M2M method is assured if our observables are of the form given by \refequation \ref{equation:observable}, where the kernels do not depend on the weights of the particles. Therefore as our observables, instead of using the mean radial velocity and velocity dispersion we use the first and second mass weighted radial velocity moments, indexed as $\vel_{r1}$ and $\vel_{r2}$. The observables are then of the form of \refequation \ref{equation:observable} with the following kernels:

\begin{align}
\label{equation:kernelVelocity} 
 K_j^{\vel_{r1}} (\vec{z}_i) &=  \delta_j^{\vel_{r1}}(\vec{z}_i) \vel_i^r \\
 \label{equation:kernelDispersion}
 K_j^{\vel_{r2}} (\vec{z}_i) &=  \delta_j^{\vel_{r2}}(\vec{z}_i) (\vel_i^r)^2 
\end{align}
where $\vel_i^r$ is the radial velocity of the particle $i$ and $\delta_j^{\vel_{r1},\vel_{r2}}$ are the field selection functions. In order to remove foreground contamination we consider only particles whose $y$ coordinate (along the GC-Sun axis) is in absolute value lower than $3.5\,\kpc$. This corresponds to the selection criterion used in the \argos survey \citep{Ness2013a} and its use here is supported by the fact that \argos data and \brava data agree with each other. We checked that the exact form of this selection function does not change significantly the kinematic observables in the inner $10\degree$.

In order to map the bias toward the near side we use the approximate luminosity function of giant stars in the bulge $\Phi(M_K) \propto 10^{0.28\, M_K}$ from WG13 where $M_K$ is the absolute magnitude in the $K$ band. In each field, stars of the \brava sample are uniformly selected between two magnitude cuts from their apparent magnitude $m_K = M_K + 5 \rm{log}(r/10 \pc) + A_K$ where $r$ is the distance to the Sun and $A_K$ is the extinction. If we consider the extinction as a foreground extinction, this uniform selection in $m_K$ is equivalent to a non-uniform selection in $r$ along the line of sight with weighting $10^{-0.28 \times 5 \, \rm{log}(r)} = r^{-1.4}$. This $r^{-1.4}$ weighting lowers the natural $r^2$ weighting due to the cone opening of the line-of-sight. We therefore adopt the following selection function:

\begin{equation}
 \delta_j^{\vel_{r1,2}}(\vec{z}_i) = 
 \begin{cases}
   r_i^{-1.4} & \text{if } i \in \text{field } j \text{ and } |y_i|<3.5\kpc,  \\
   0 & \text{otherwise.}
  \end{cases}
\end{equation}

In order to compare the model observables to the target data, we have to weigh the target data by the expected mass in each field. As the stellar mass in each field is unknown, we use the model mass, and update this weighting several times during the fit (see {\refsection \ref{section:M2Mparametrization}}).

\subsubsection{Proper motions}
\label{section:properMotionObservables}
In addition to the \brava data, we use proper motion data from \citet{Rattenbury2007}. These authors computed proper motion dispersions, $\sigma_{l,b}$, in the $l$ and $b$ directions for a large number of bulge RCGs in $45$ bulge fields from proper motion measurements for stars in the \textsc{ogle-ii} survey. We use these proper motions as a check of our modelling, comparing the data to our model predictions, without fitting them. \citet{Rattenbury2007} selected bulge RCGs from their colours and magnitudes, and excluded all stars with total proper motion larger than $10 \masyr$. The bulge membership of model particles is evaluated using the same criterion as for the \brava observables, i.e. $|y| \leq 3.5\kpc$. Similarly to the \brava observables, our proper motion observables are the first and second velocity moments in the $l$ and $b$ direction, indexed by $\vel_{l1,2}$ and $\vel_{b1,2}$. The kernels of the first and second velocity moments in the $l$ and $b$ directions are similar to those of \refequations \ref{equation:kernelVelocity} and \ref{equation:kernelDispersion}, replacing $\vel_i^r$ by the velocity of particle $i$ in the $l$ and $b$ direction, expressed in $\masyr$.

Our selection function when evaluating the model proper motion observables is then given by:

\begin{equation}
 \delta_j^{\vel_{l1,2}, \vel_{b1,2} }(\vec{z}_i) = 
 \begin{cases}
   1 & \text{if } i \in  \text{field } j\text{, } |y_i|<3.5\kpc\\
   &\text{and } \sqrt{\vel_l^2 + \vel_b^2} \leq 10 \masyr\\
   0 & \text{otherwise}
  \end{cases}
\end{equation}

\subsection{Dynamical velocity scaling}
\label{section:dynamicalScaling}

Our models are evolved in a system of internal units where the length unit, velocity unit and gravitational constant are set to unity. When comparing model to data we scale the models to physical units using the length of the bar (see \refsection \ref{section:geometryAndScaling}), the gravitational constant and a velocity scaling. This velocity scaling is first fixed to some reference value using the circular velocity at the radius of the Sun from \citet{Bovy2012} ($218 \kms$). Keeping the velocity scaling fixed to this value would be inappropriate, however: our models are constrained in the inner part by the \brava data and the 3D density but no effort has been made to make the model rotation curves match the MW rotation curve at larger radii. Therefore the velocity scaling given by the circular velocity at the radius of the Sun is not relevant in the bulge. Hence we determine the velocity scaling directly from the \brava data using a variant of the method presented by \citet{DeLorenzi2008}. In \refequations \ref{equation:kernelVelocity} and \ref{equation:kernelDispersion}  we replace the radial velocity of a particle $\vel_i^r$ by $\gamma\,\vel_i^r$, where $\vel_i^r$ is now the radial velocity of particle $i$ expressed in physical units using the reference velocity scaling, and $\gamma$ is a numerical factor initially set to one. Through {\refequations \ref{equation:observable}} and {\ref{equation:delta}} the total $\chi^2$ ({\refequation \ref{equation:chi2}}) and the profit function $F$ ({\refequation \ref{equation:profit}}) now depend on $\gamma$. To find the maximum of $F$ with respect to $\gamma$, we use the following evolution equation for $\gamma$, similar to the force-of-change equation:
\begin{equation}
 \label{equation:FOCgamma}
 \frac{d\gamma}{dt} = - \eta \gamma \frac{\partial \chi^2}{\partial \gamma}
\end{equation}
where $\eta$ sets the magnitude of the force-of-change applying on $\gamma$. The velocity scaling plays a role for the kinematic observables only, so the derivative of $\chi^2$ is given by
\begin{equation}
 \frac{\partial \chi^2}{\partial \gamma} = 2 \sum_{j} \lambda_{\vel_{r1}} \Delta_j^{\vel_{r1}} \frac{\partial \Delta_j^{\vel_{r1}}}{\partial \gamma} + \lambda_{\vel_{r2}} \Delta_j^{\vel_{r2}} \frac{\partial \Delta_j^{\vel_{r2}}}{\partial \gamma}
\end{equation}
where the derivatives of the $\Delta_j^{\vel_{r1,2}}$ are given by the following equations:
\begin{align}
 \frac{\partial \Delta_j^{\vel_{r1}}}{\partial \gamma} &= \frac{y_j^{\vel_{r1}}}{\gamma\, \sigma(Y_j^{\vel_{r1}})} \\
 \frac{\partial \Delta_j^{\vel_{r2}}}{\partial \gamma} &= \frac{2 \, y_j^{\vel_{r2}}}{\gamma\, \sigma(Y_j^{\vel_{r2}})}
\end{align}

The value of $\eta$ should be large enough to ensure the convergence of $\gamma$ during the fit, but the time-scale for its evolution must be longer than the temporal smoothing time-scale. We fixed $\eta$ to $0.4$ from estimating the magnitude of the right hand side of {\refequation \ref{equation:FOCgamma}}. In all our fits, $\gamma$ converges to some final value which we use to convert internal units to physical units. All velocities and proper motions are scaled by $\gamma$ with respect to their value with the reference scaling and all masses are scaled by $\gamma^2$.

\subsection{Potential and model integration}
\label{section:potential}
Stellar and dark matter particles are all integrated in their combined potential that comes initially from our models of barred discs evolved in dark matter haloes (see {\refsection \ref{section:particleModels}}). Only stellar particles are used as M2M particles, whose masses are both used as gravitational masses and as M2M weights. During the \nmagic fits, the gravitational potential changes as the particle weights in the bulge adapt to match the target density (including the $\F$ factor) and the kinematic constraints. To take this into account, the potential of the stellar and halo particles is recomputed from time to time during the fit. This ensures that the new weights of the disc particles can influence the kinematics of the model and that the final converged model evolves in its own gravity.  In between potential updates, the particles are integrated in a frozen rotating potential. The potential always rotates at the constant pattern speed $\Omega_{\rm p}$ of the initial model, in internal units, and the centre of mass as well as the rotation axis of the potential are kept fixed during the complete \nmagic fit.  Each potential update corresponds to an update of the orbit library.

In all our fits the potential is recomputed 10 times during the weight evolution, which we found is sufficient to ensure the self-gravity of the converged model as well as smooth updates of the orbit library. The potential is computed using the 3D polar grid code from \citet{Sellwood1997}. The particles are integrated with a drift-kick-drift adaptive leap-frog algorithm. The integration scheme is such that over a typical fit integration time in a fixed potential, the Jacobi energy is conserved to a level of $10^{-3}$ or better.

\subsection{\nmagic parametrization}
\label{section:M2Mparametrization}

\subsubsection{Fitting procedure}
A typical \nmagic fit consists of the three following phases. First we evolve the particles $\rm{T_{smooth}}$ iterations during which we compute the model observables in order to initialize the temporally smoothed observables. Then we integrate the model for $\rm{T_{M2M}}$ iterations, changing the weights of the particles according to \refequation \ref{equation:FOC2} while updating the potential regularly. All observables are matched to the data at the same time. After a last potential update we finally relax the model for $\rm{T_{relax}}$ iterations without changing the particle weights. This last phase is important to avoid over-fitting the data and obtain realistic models. The usually slight $\chi^2$ increase during this relaxation reveals how much the model was forced to fit the data by \refequation \ref{equation:FOC2}. At the end of the run, we also checked the convergence of the particle weights, using the convergence criterion detailed in \citet{Long2010}. A particle weight is considered to have converged if its maximum relative deviation from its mean value over some period of time is smaller than some threshold. Assuming a period of time corresponding to four circular orbits at $2\kpc$ and a threshold of $10\%$, about $97\%$ of the particles weights converge in a typical \nmagic fit.

\subsubsection{Parameter values and time-scales}
The parameter values we use are shown in \reftable \ref{table:fitParameters}. The iteration step of the M2M procedure $dt$ is fixed to one thousandth of the time needed to complete a circular orbit of radius $2\kpc$. All models are integrated for a constant number of $dt$ so the number of bar rotations during the weight adaptation phase varies from model to model, ranging between $25$ for model M80 and $35$ for model M90. This corresponds to a physical integration time between $6$ and $7\, \Gyr$, once scaled to physical units using the velocity scaling determined dynamically by \nmagic.
 $1/(\alpha * dt)$ is the time-scale of the temporal smoothing expressed in terms of number of M2M iterations, and $\varepsilon/ w_0$ is the magnitude of the force-of-change in \refequation \ref{equation:FOC2}, where $w_0$ is the initial weight of a stellar particle.

\begin{table}
\caption{Typical set of \nmagic parameters. $\rm{T_{smooth}}$, $\rm{T_{M2M}}$ and $\rm{T_{relax}}$ are the number of iterations for the 3 phases of a \nmagic fit. $\rm{T_{\Phi}}$ and $\rm{T_{mass}}$ are the number of iterations between two potential updates and updates of the mass weighting for the kinematic observables. $1/(\alpha * dt)$ is the time-scale of the temporal smoothing in number of iterations. $\varepsilon/w_0$ is the magnitude of the force of change and $w_0$ is the initial weight of the stellar particles. The $\lambda$ parameters are described in \refsection \ref{section:M2Mlambda}.}
  \centering
  \begin{tabular}{ccccc}
    $\rm{T_{smooth}}\,[\iteration]$ & $\rm{T_{M2M}}\,[\iteration]$ & $\rm{T_{relax}}\,[\iteration]$ & $\rm{T_{\Phi}}\,[\iteration]$ & $\rm{T_{mass}}\,[\iteration]$ \\
    \hline\hline
    $10^4$ & $10^5$ & $2 \times 10^4$ & $10^4$ & $10^4$  \\
    \\

    $1/(\alpha * dt)\,[\iteration]$  & $\varepsilon / w_0 [\iu^{-1}]$& $\lambda_{\rm{d}}$ & $\lambda_{\vel_{r1}, \vel_{r2}}$ &\\
    \hline\hline
    $2.5\times10^3$  & $0.04$ & $1$ & $25$ &\\

  \end{tabular} 
  \label{table:fitParameters}
\end{table}

\subsubsection{$\lambda_k$ parameters and regularization}
\label{section:M2Mlambda}

The different set of observables, here RCGs 3D density and \brava data are weighted by the $\lambda_k$ parameters in the profit function. The force-of-change (\refequation \ref{equation:FOC2}) already takes into account observational errors so in theory these $\lambda_k$ should all be set to $1$, in order to really minimize the total $\chi^2$. In practice, experiments showed that with all $\lambda_k = 1$ the model ignores completely the kinematic constraints. This is mostly caused by the very large number of density constraints ($26880$) with respect to kinematic constraints ($164$). As we do want to fit the \brava data, we increase the weighting of \brava constraints with respect to the RCGs density using $\lambda_{\vel_{r1}} / \lambda_{\rm{d}} = \lambda_{\vel_{r2}} / \lambda_{\rm{d}} = 25$, even if then we no longer strictly minimize the total $\chi^2$. These $\lambda_k$ were determined using the distribution of the force-of-change contribution of each set of observables, such that the mean force-of-change due to the density observables should be equal to the mean force-of-change due to the \brava observables. This causes the \brava data to be reasonably fitted without being over-fitted. A strong over-fitting would lead to a significant increase of $\chi^2$ during the relaxation step at the end of the fit. 

We found that our models were smooth enough without using any entropy smoothing. This is a consequence of the very dense constraint from the 3D density data. Hence we chose to set $\mu = 0$ in \refequation \ref{equation:FOC2} for all our models.

\section{Dynamical models of the MW}
\label{section:DynamicalModels}

In this section we fit our initial models to the \brava data and the 3D RCGs density using \nmagic. The models differ by their dark matter fraction in the inner part and by their dimensionless corotation radius $\R$ (see \refsection \ref{section:geometryAndScaling}). In \refsection \ref{section:DynamicalModelsBestF}, we determine the best bulge stellar mass $\F$ for each model, using the full \nmagic modelling procedure described above, and in \refsection \ref{section:DynamicalModelsBestModels}, we compare all models with their respective best $\F$. In \refsection \ref{section:DynamicalModelsPatternSpeed}, we constrain the pattern speed of the MW bar from this modelling, and in \refsection \ref{section:PM} we compare the models with available proper motion data.

\subsection{Determination of the stellar mass in the bulge}
\label{section:DynamicalModelsBestF}

\begin{figure*} 
  \includegraphics{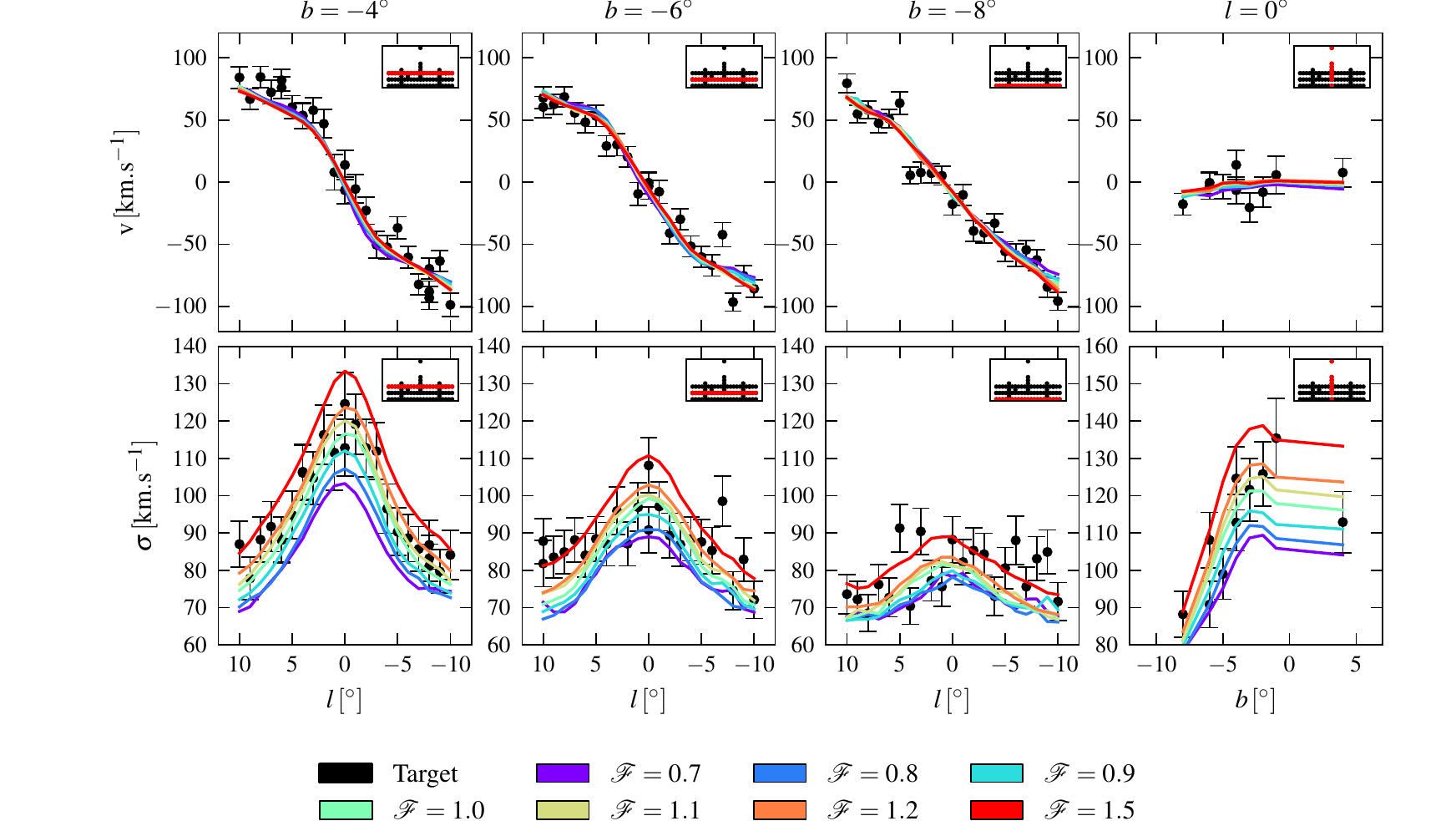}
  \caption{Comparison of the \brava data to the fitted model M85, for different values of the stellar mass parameter $\F$. The upper (lower) plots show the \brava mean radial velocity (velocity dispersion) profiles for $b = -4\degree$, $b = -6\degree$, $b = -8\degree$ and along the minor axis $l = 0\degree$. The different colours indicate different values of $\F$ as stated in the legend. The influence of $\F$ on the kinematics is clearly visible in the dispersion plots. The small inserts on the top right of each plot show the \brava fields in Galactic coordinates and highlight the fields shown in the corresponding plot.}
    \label{fig:allMasses}
\end{figure*}

In \refsection \ref{section:M2MDensityObservable} we parametrized the target density observables using a free numerical factor $\F$. The value of $\F$ directly sets the stellar mass of the bulge in internal units for each model. We find that $\F$ has a strong influence on the velocity dispersions but only a very slight influence on the mean velocities. Empirically, the shape of the mean velocity profile in our models is fixed by the density distribution. The velocity  amplitude in internal units is then fixed by the pattern speed {\citep[\refequation~8 of][]{Debattista2002}}. The larger kinetic energy in models with larger bulge mass $\F$ can therefore only be put into the velocity dispersion. This is shown in \reffigure \ref{fig:allMasses} where the kinematic observables of model M85 are plotted along with the \brava data. In this figure the different coloured lines show the kinematic profiles obtained after fit of the density normalized with different values of $\F$. The upper (lower) plots show mean radial velocity (velocity dispersion) profiles for three different latitudes as well as along the minor axis. For the comparison, the model observables in {\reffigure \ref{fig:allMasses}} have all been plotted for the same scaling constant so as to match only the mean velocity data. As $\F$ has nearly no influence on the velocity, this better highlights the effect of $\F$ on the dispersion.

\begin{figure} 
  \includegraphics{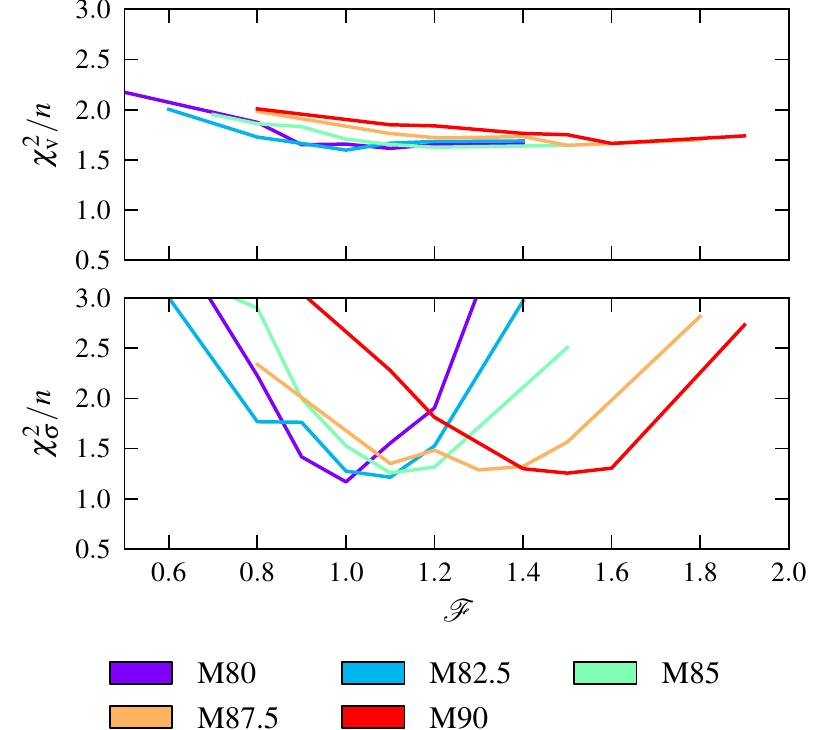}
  \caption{$\chi^2$ per data point of the velocity observables (\emph{upper plot}) and velocity dispersion observables (\emph{lower plot}) as a function of the bulge mass factor $\F$. Each colour represents a different initial model from \refsection \ref{section:particleModels} as shown in the legend. In all cases, a best value of $\F$ is clearly visible from the velocity dispersion $\chi^2$-curves.}
    \label{fig:chi2k}
\end{figure}

\reffigure \ref{fig:allMasses} illustrates how more massive bulges lead to higher radial velocity dispersions. By finding the value of $\F$ which gives the best agreement with the data, we can recover the stellar mass of the $\bb$ for each model. \reffigure \ref{fig:chi2k} shows the $\chi^2$ per data point of the velocity and velocity dispersion for all models plotted in \reffigure \ref{fig:allMasses}, versus $\F$. As expected, a clear minimum is present in the dispersion plot and provides us with a best value of $\F$ for each model. Best values of $\F$ are given for all models in the first column of \reftable \ref{table:chi2Fitting}. With the optimal velocity scaling determined in these \nmagic fits, these $\F$ values can be converted into physical values of stellar mass in the $\bb$. We find values of $1.25\times\SM$ for model M80 to $1.6\times\SM$ for model M90; see also \refsection \ref{section:mass}.  

$\F$ was sampled on a regular grid with spacing $0.1$ and is therefore determined with an accuracy of $0.05$. Due to the rescaling during the \nmagic fits, an uncertainty of $0.05$ in $\F$ typically corresponds to a change of less than $0.025\times\SM$ in the final stellar mass and less than $0.01\times\SM$ in total mass (stellar and dark matter in the $\bb$).

\subsection{Best dynamical models of the Milky Way bulge}
\label{section:DynamicalModelsBestModels}

\begin{figure}
  \includegraphics{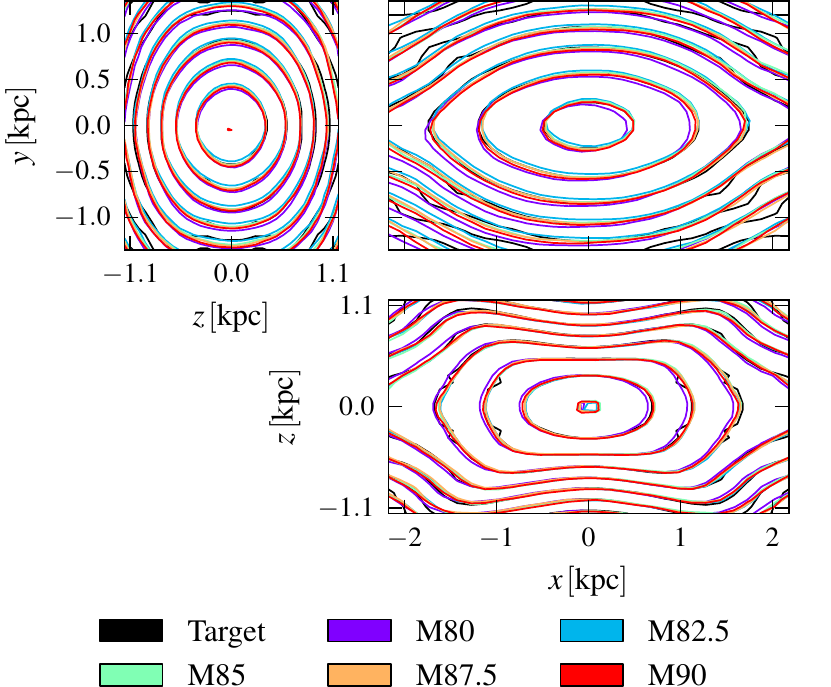}
  \caption{Contour plot of the projected 3D density in the bulge of our best dynamical mass models compared to the measured RCGs density from \citet{Wegg2013}. The projections are the same as in \reffigure \ref{fig:RCGDensity} and the contours are spaced by a third of a magnitude. In all cases the density is well fitted.}
   \label{fig:fittedDensity}
\end{figure}

Now we compare our five best dynamical models M80-M90 with different dark matter haloes, obtained after \nmagic fit to the data for their respective best value of $\F$. For all models the density and its peanut shape is well fitted as shown in \reffigure \ref{fig:fittedDensity}. This figure compares the contours of the three principle axis projections of the density in the bulge after fitting with the target RCGs density. Each colour line represents one of our best mass models and the black line is the projection of the target density. 

\begin{figure*}
  \includegraphics{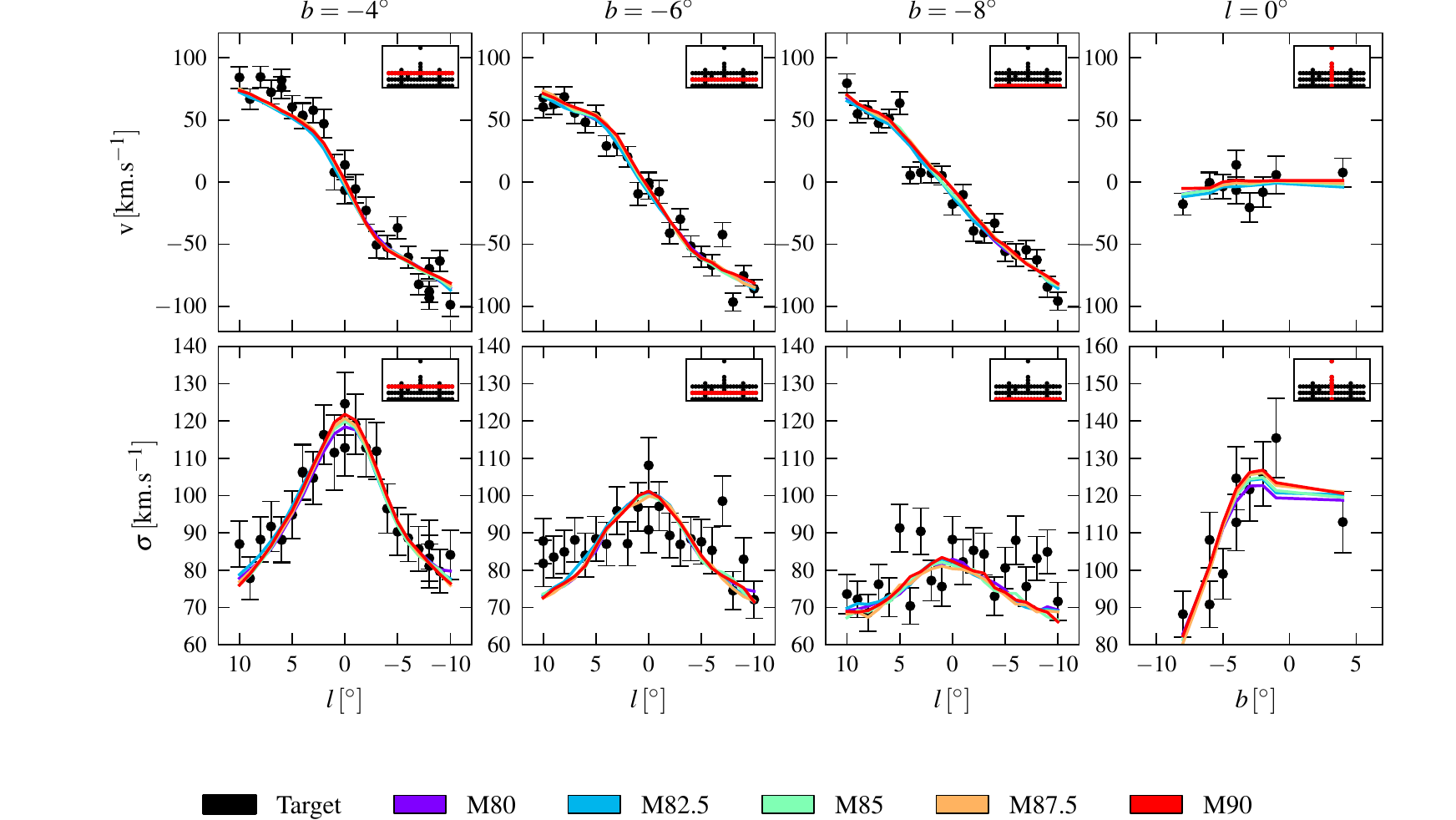}
  \caption{Velocity and velocity dispersion profiles for our five best dynamical models of the Milky Way with different dark matter haloes. The plotting conventions are the same as in \reffigure \ref{fig:allMasses} except that here the colours indicate different models as stated in the legend.}
   \label{fig:fittedKinematics}
\end{figure*}

The radial velocity and dispersion profiles of the same best $\F$ models are plotted in \reffigure \ref{fig:fittedKinematics}. Except for small differences in the shapes of the dispersion curves, they all look very similar and provide a good fit to the data. This similarity is explained by the fact that the shape of the velocity profile is mostly fixed by the shape of the density and its magnitude is adapted to the data by the floating velocity scaling. The magnitude of the velocity dispersion profiles can then be adapted independently by the $\F$ parameter.

\begin{table}
\caption{Best values of $\F$ and $\chi^2$ per data point of the density observables ($\chi^2_{\rm{d}}/n_{\rm{d}}$), total kinematic observables ($\chi^2_{\rm{k}}/n_{\rm{k}}$), velocity only ($\chi^2_{\vel}/n_{\vel}$) and dispersion only ($\chi^2_{\sigma}/n_{\sigma}$) for our five best dynamical mass models. The last column shows the reduced $\lambda_k$ weighted $\chi^2$ actually minimized by \nmagic.}
\label{table:chi2Fitting}
  \centering
  \begin{tabular}{l|cccccc}
    Model & $\F$ & $\chi^2_{\rm{d}}/n_{\rm{d}}$& $\chi^2_{\rm{k}}/n_{\rm{k}}$ & $\chi^2_{\vel}/n_{\vel}$& $\chi^2_{\sigma}/n_{\sigma}$ & $\chi^2_{\rm{tot}}/n_{\rm{tot}}$\\
    \hline\hline
M80 & 1.0 & 0.42 & 1.45 & 1.67 & 1.24 & 0.56 \\
M82.5 & 1.1 & 0.32 & 1.46 & 1.68 & 1.24 & 0.47 \\
M85 & 1.1 & 0.24 & 1.46 & 1.65 & 1.27 & 0.40 \\
M87.5 & 1.3 & 0.23 & 1.52 & 1.73 & 1.31 & 0.40 \\
M90 & 1.5 & 0.30 & 1.52 & 1.75 & 1.29 & 0.46 \\
  \end{tabular} 
\end{table}

More quantitatively, the $\chi^2$ per data point is shown in \reftable \ref{table:chi2Fitting} separately for the density, the total kinematics, the velocity only and the velocity dispersion only. Values in this table are not weighted by the corresponding $\lambda_k$ parameters. The total $\lambda_k$-weighted chi square $\chi^2_{\rm{tot}}/n_{\rm{tot}} = \sum \lambda_k \chi^2_k / \sum \lambda_k n_k$ actually minimized by \nmagic is given in the last column of \reftable \ref{table:chi2Fitting}.

All our models provide good fits of the RCGs density with a final $\chi^2_{\rm{d}}/n_{\rm{d}}$ from $0.23$ to $0.42$. As the errors in the density are systematic, one should not over-interpret these $\chi^2$ values. The kinematics are also well fitted for all models with the model dispersions marginally steeper than the data for large $\vert l\vert$ and $\vert b \vert$, and $\chi^2_{\rm{k}}/n_{\rm{k}}$ ranging from $1.45$ for model M80 to $1.52$ for model M90. This kinematic $\chi^2$ is significantly better than that obtained by \citet{Long2013}. Given the scatter visible in the kinematic data, these models are good candidates to represent the MW bulge even if the kinematic $\chi^2_{\rm{k}}$ is about $1.5$.

Because the different models all give a reasonable fit, we do not rule out some of them based on simple $\chi^2$ considerations. One has to be careful when drawing conclusions comparing $\chi^2$ values on typical M2M problems. Indeed, as shown by \citet{Morganti2013} the common practice to use  $\Delta \chi^2 = \chi^2 - \chi^2_\textrm{min}$ and $\chi^2$ statistics to evaluate confidence levels on $\chi^2_\textrm{min}$ is not appropriate for their M2M results. The $\Delta \chi^2$ analysis requires a positive number of degrees of freedom, while it is usually negative in M2M problems because the particles are vastly more numerous than the data constraints. Hence, the $\chi^2/n$ given in \reftable \ref{table:chi2Fitting} are only an indication of the distance between model and data. These values are also strongly influenced by a few data points which appear to be possible outliers. This is the case for example in the fields at $(l,b) = (-7\degree,-6\degree); (-5\degree,-4\degree); (-4\degree,-8\degree)$. Removing these possible outliers reduces the absolute value of the kinematic $\chi^2_k/n_k$ by about $0.2$. It does not however change our best value of $\F$.

\begin{figure}
  \includegraphics{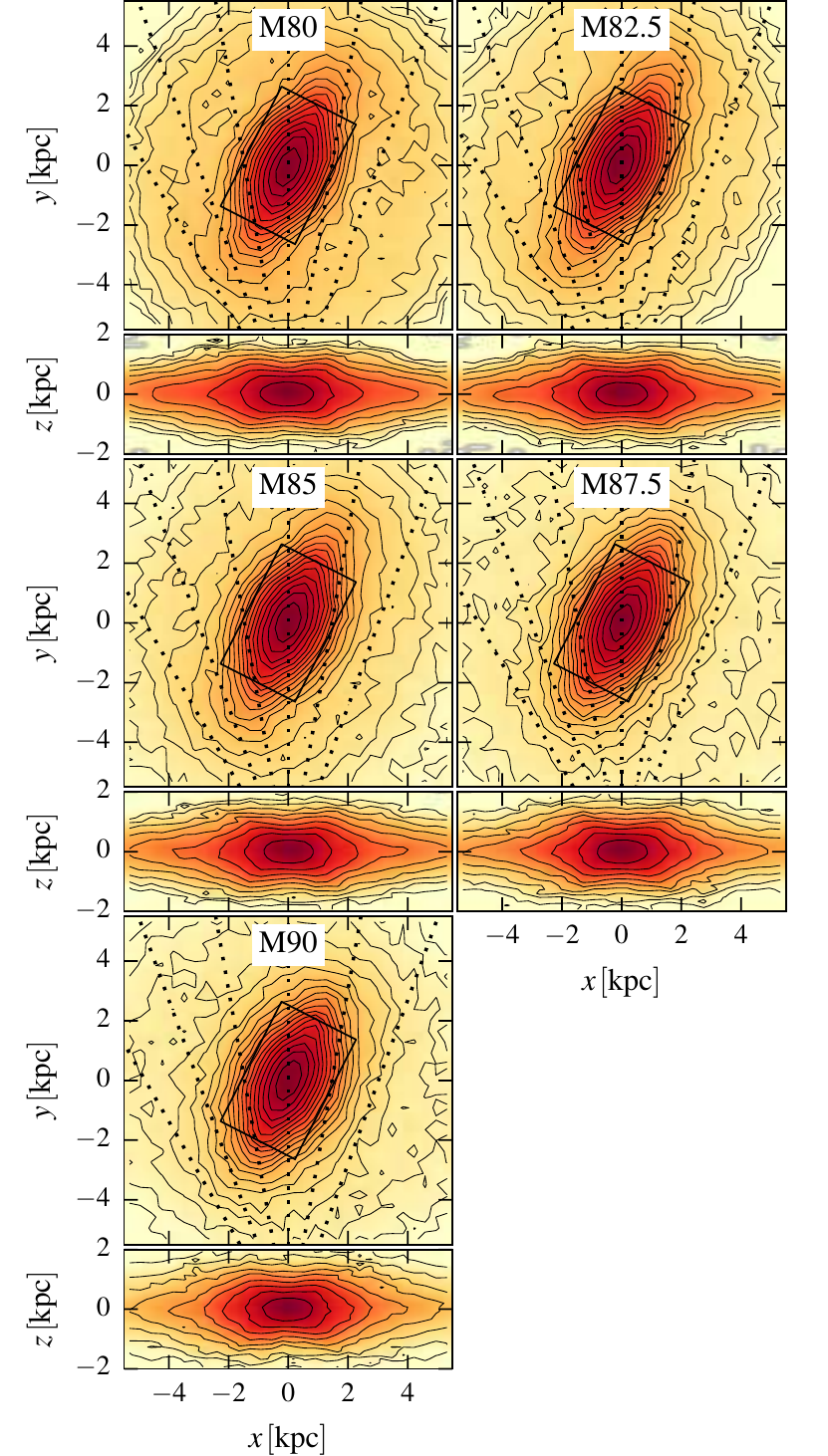}
  \caption{Face-on and side-on views of our best dynamical mass models after \nmagic fit. Plotting conventions are the same as in \reffigure \ref{fig:modelsPlots}.}
  \label{fig:fittedModels}
\end{figure}

The face-on and side-on projections of our final best mass models are plotted \reffigure \ref{fig:fittedModels}. Even though not enforced by \nmagic, the long bar component is still there in the fitted models. Its presence indicates that the gravitational potential updates performed during the fit were smooth enough to keep long bar particles on bar orbits.

\subsection{The pattern speed}
\label{section:DynamicalModelsPatternSpeed}

The pattern speeds of our models are converted to physical units using the velocity scaling determined by \nmagic in order to fit the \brava data. For models M80, M82.5, M85, M87.5 and M90 we get the values $\Omega_{\rm p}\,[\kmskpc] = 24.7, \, 25.7, \, 27.7, \, 29.0, \, 28.8$. This shows that the pattern speed of the MW bar and bulge in absolute units, as determined by the combination of the RCGs density and the \brava data, is between $25$ and $30\kmskpc$. This is slightly lower than the value of $\Omega_{\rm p} \sim 30-40\kmskpc$ determined by \citet{Long2013}, also from the \brava data. The comparison with other determinations in the literature is discussed in \refsection \ref{section:DiscussionPatternSpeed}.

\begin{figure}
  \includegraphics{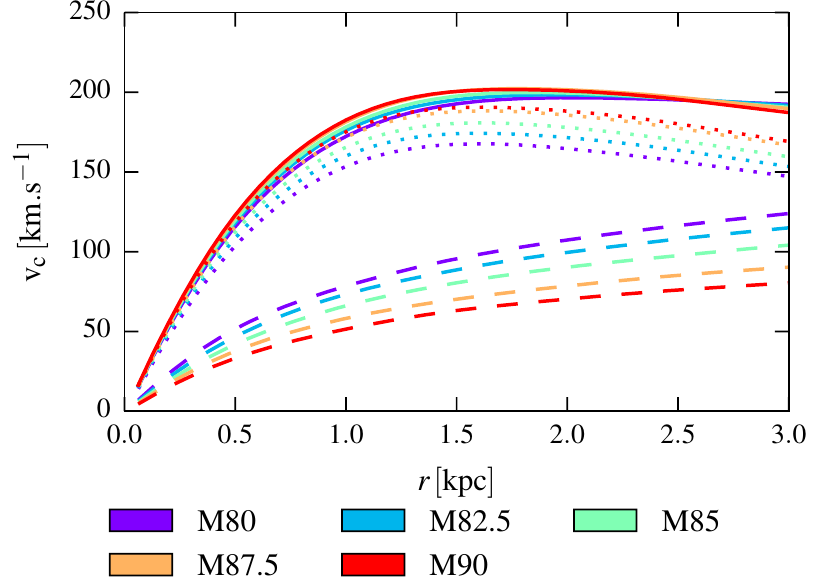}
  \caption{Circular velocity curves obtained from the azimuthally averaged potential of our best dynamical mass models after \nmagic fit. The solid lines display the total rotation curves while dashed and dotted lines display the halo and disc contributions, respectively.}
  \label{fig:finalRotationCurve}
\end{figure}

Our initial models were constructed with different dark matter haloes, and the bars formed in these models had different corotation radii and $\R$ values $\R=1.1-1.8$ based on their individual rotation curves (\reffigure \ref{fig:circularVelocity}). After the rescaling during the \nmagic fit, the rotation curves of all models are essentially identical in the inner $3\kpc$, such that these models all provide equally good fits to the RCGs density and \brava data. This is displayed in \reffigure \ref{fig:finalRotationCurve} which shows the azimuthally averaged rotation curves of our best mass models, together with their disc and halo contributions. The models' scaled outer rotation curves are different, however, consistent with the different $\R$ values. This is not expected to influence the bulge dynamics, because very few stars from $4\kpc$ and beyond will reach the \brava bulge fields along their orbits.

Because no effort has been made to match the rotation curve of the MW at large radii, the models' $\R$ ratio does not correspond to that of the MW. There are several ways in which the outer rotation curve of the MW could have changed after the bar and bulge formed, e.g., by later growth of the disc. Therefore, in order to estimate the corotation radius and $\R$ of the Galaxy corresponding to $\Omega_{\rm p} \sim 25-30\kmskpc$ we need to use additional data. Assuming the composite rotation curve of \citet{Sofue2009} rescaled to $(R_0, V_0) = (8.3\kpc, 218\kms)$, this range of pattern speed would result in a corotation radius between $7.2$ and $8.4\kpc$, which implies $\R$ between $1.5$ and $1.8$. The MW would then belong to the so-called slow rotators. This result is discussed in more detail in \refsection \ref{section:DiscussionPatternSpeed}.

\subsection{Proper motions}
\label{section:PM}

\begin{figure*}
  \includegraphics{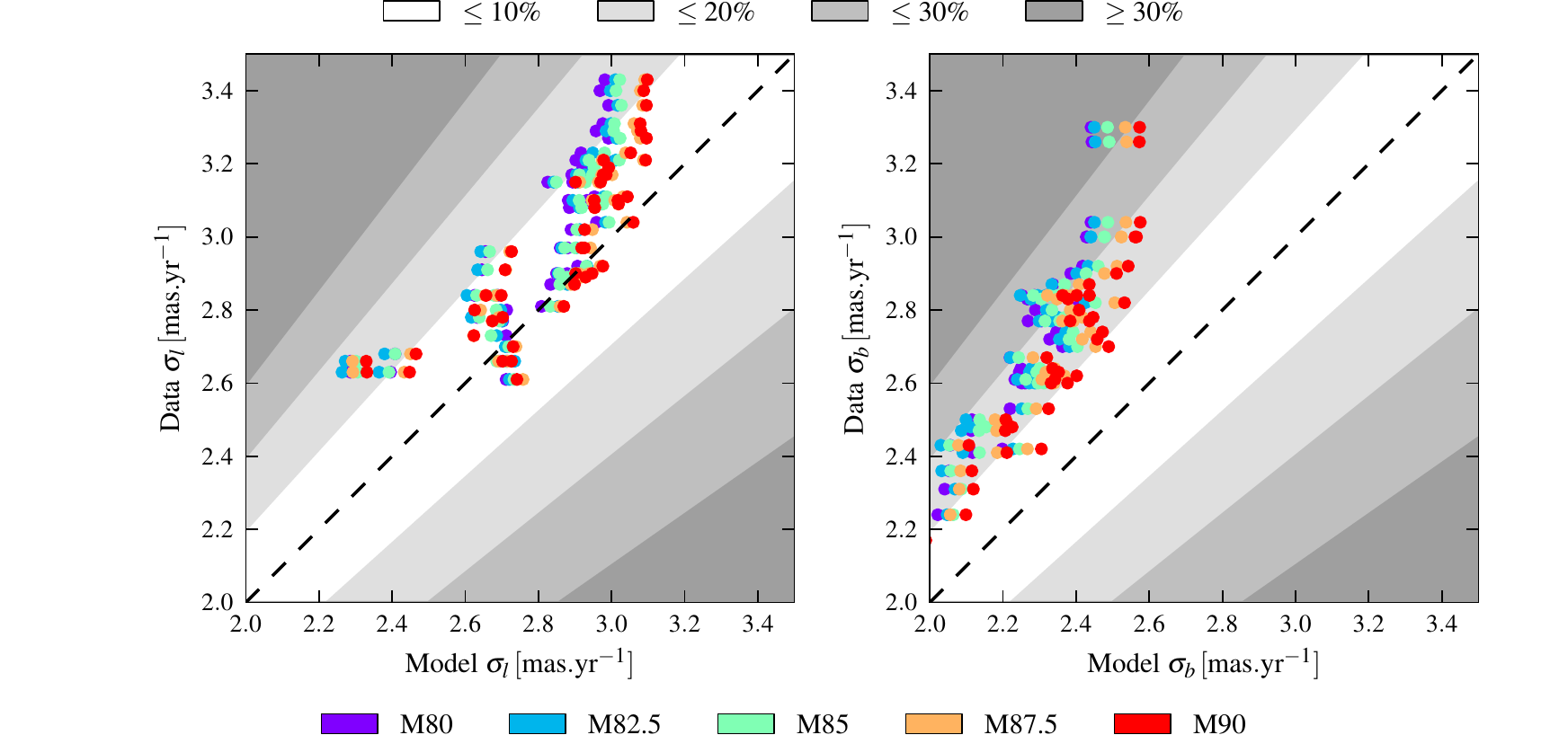}
  \caption{Comparison between model and data for proper motion dispersions in the $l$ direction (left-hand plot) and in the $b$ direction (right-hand plot) for all bulge fields from \citet{Rattenbury2007}. The different colours refer to the models as stated in the legend. The shaded regions indicate the relative error of model predictions with respect to the data, from $\leq 10\%$ in white to $\leq 20\%$, $\leq 30\%$ and $\geq 30\%$ in dark grey.}
  \label{fig:PM}
\end{figure*}

As an independent check we predict proper motion dispersions $\sigma_{l,b}$ in the $l$ and $b$ direction for our best dynamical models as explained in {\refsection \ref{section:properMotionObservables}}, and compare them to the data from \citet{Rattenbury2007}. A comparison model/data is shown in \reffigure \ref{fig:PM} for $\sigma_l$ (left-hand plot) and $\sigma_b$ (right-hand plot). The different colours refer to the five best dynamical models, and the shaded regions display different levels of relative error of the model with respect to the data. Data error bars are not plotted: errors given by \citet{Rattenbury2007} are only statistical errors at the level of $1\%$ or better. However, reproducing their derivation of $\sigma_{l,b}$ from the original motions of individual stars in a couple of fields, it seems to us that systematic errors dominate. These systematic effects are due to the selection thresholds and have a typical magnitude of $10\%$, which is indicated in \reffigure \ref{fig:PM} by the white band.

Our models provide good proper motion predictions in the $l$ direction, being mostly inside the $10\%$ limit. Proper motions in the $b$ direction are slightly worse, with model values mostly $10\%$ to $20\%$ lower than the data. This hints at additional systematic errors, either in the models or in the data. $\sigma_b$ is directly related to the vertical derivative of the potential and therefore to the mass concentration toward the midplane. This is apparent in \reffigure \ref{fig:PM} where models with a more massive stellar bulge component have larger $\sigma_b$. 
There is only limited scope for increasing $\sigma_b$ in the models as will be discussed in \refsection \ref{section:DiscussionDarkHalo}. \citet{Rattenbury2007} noted that deviations of proper motion dispersions between adjacent fields could be as high as $0.2\masyr$ and concluded that some small systematic effect could indeed be present in the data.
All together our models predict reasonable proper motion dispersions, indicating that the dynamics of the GB is quite constrained by the combination of the \brava data and the RCGs density.

\section{Mass of the Galactic bulge}
\label{section:mass}
\subsection{Evaluation of the total mass}
\label{section:MassEvaluation}

\begin{figure}
  \includegraphics{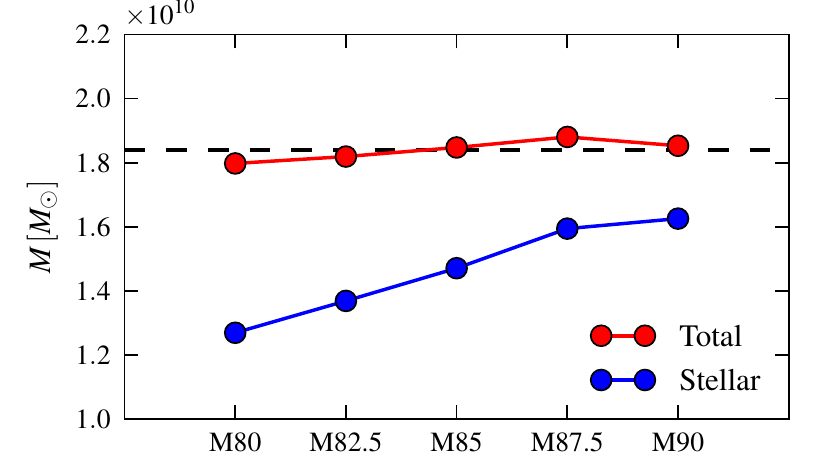}
  \caption{Mass of the bulge-in-box for our five best mass models in different dark matter haloes. The blue curve refers to the stellar mass while the red curve refers to the total mass.}
  \label{fig:mass}
\end{figure}

The stellar mass in the bulge-in-box ($\bb$, the inner $\pm 2.2 \times \pm 1.4 \times \pm 1.2\kpc$ of the GB) is determined  by the value of $\F$ derived in the previous section together with the velocity scaling found by \nmagic during the fit. The stellar mass recovered this way differs from model to model, ranging from about $1.25\times\SM$ for model M80 to $1.6\times\SM$ for model M90. This is expected given the fact that our models have different dark matter masses. More massive haloes (like M80) can build the \brava dispersion with relatively low stellar mass while low mass haloes need more stellar mass. All our models are good fits to the data, so purely from the \brava data we cannot infer the stellar mass of the bulge accurately. However, our modelling gives us a very good estimate of the \emph{total} mass of the bulge-in-box. \reffigure \ref{fig:mass} shows the stellar mass (blue points) and total mass (red points) of the $\bb$ for all models. We can see that our estimates of the total mass are quite constant along our range of model haloes. Altogether we evaluate the total mass of the $\bb$ to be $1.84 \pm 0.07 \times \SM$. The errors quoted here are combined statistical and systematic whose determination is discussed below.

To estimate statistical errors, we use a $\Delta\chi^2$ analysis, based on two approaches. First, we regard the velocity dispersion profiles for a given model with different $\F$ parameters as a one-parameter family of curves matched to the 82 \brava velocity dispersions, in which case the appropriate $\Delta\chi^2=1$. This leads to an average uncertainty for the models of $\Delta\F=0.04$, $\Delta \Ms=0.025 \times \SM$, $\Delta \Mtot=0.028 \times \SM$. Secondly, we consider the kinematic $\chi_k^2/n_k$-values of the best-fitting models from {\reftable \ref{table:chi2Fitting}} for the different dark matter haloes as the combination of a systematic variation modelled as linear, plus a fluctuating component which has a root mean square of rms$(\chi_k^2)=0.0119\times 164= 1.95$. We take this as an approximation of the scatter in $\chi_k^2$ at minimum, which according to the simulations of {\citet{Morganti2013}} can be taken as a proxy for the $\Delta\chi^2$ to be used for estimating the accuracy with which the mass can be determined from the data using \nmagic modelling. Applying this to the combined $\chi_k^2$ curve derived from {\reffigure \ref{fig:chi2k}} results in estimated uncertainties of $\Delta\F=0.05$, $\Delta \Ms=0.033 \times \SM$, $\Delta \Mtot=0.036 \times \SM$. Based on both methods, we estimate the statistical uncertainty in the stellar and total mass measurement for the $\bb$ as $\Delta \Ms=0.03  \times \SM$, $\Delta \Mtot=0.03 \times \SM$.

\subsection{Evaluation of systematics}
\label{section:Systematics}
Our modelling relies on some assumptions whose influence on the derived bulge mass we now investigate. Here we describe variations of our four main assumptions: the midplane extrapolation, the length scaling, the snapshot selection and the bar angle. We show that the effect of these assumptions on the estimate of the mass of the bulge is small. These test variations are then used to set the error on the mass measurement previously quoted.

\subsubsection{Midplane extrapolation}
\label{section:variantExtrapolation}

\begin{figure}
  \includegraphics{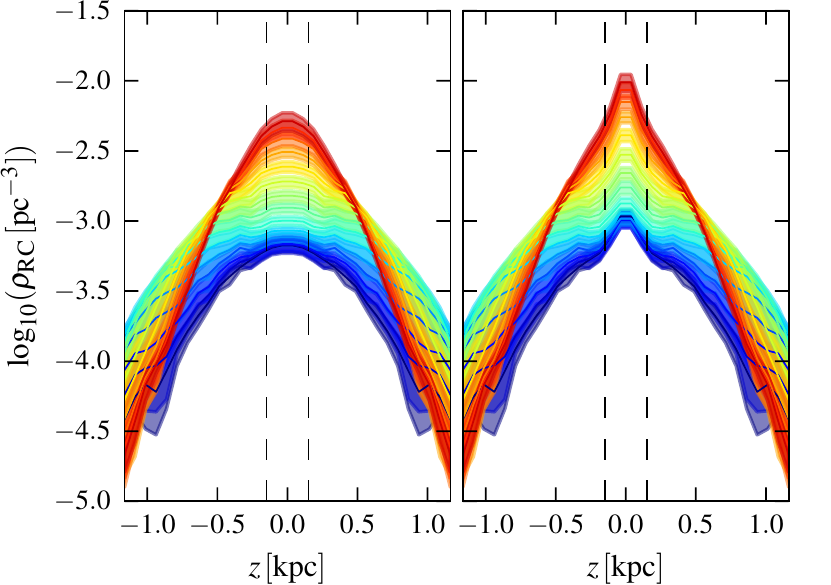}
  \caption{Vertical density profiles along the major axis of the bar for the fiducial extrapolation of the density to the midplane (left plot) and the variant extrapolation (right plot). The different colours indicate the coordinate along the major axis, from red in the centre to blue at the edge of the map at $2.2 \kpc$. The colour shaded region depicts the errors in the density from \citet{Wegg2013} and the dashed lines show the $\pm 150 \pc$ region where the original density map from \citet{Wegg2013} has been extrapolated.}
  \label{fig:alternativeExtrapolation}
\end{figure}

In order to quantify the uncertainty introduced by the assumed shape of the extrapolation in the midplane (see \refsection \ref{section:M2MDensityObservable}), we do the same study as detailed in \refsection \ref{section:DynamicalModels} with a variant extrapolation, more concentrated in the midplane. 

In the fiducial case, we use the best $\rm{sech}^2$ fit to fill in the midplane. For our variant, we perform an exponential extrapolation down to the midplane, with scale height $h$ varying as a function of the in plane coordinates $(x',y')$ (see \refsection \ref{section:geometryAndScaling}). We assume that $h(x', y')$ is proportional to the scale height of the best $\rm{sech}^2$ fit at large $z$, normalized with numerical value fixed such that $h(0,0)$ is equal to $1\degree \sim 140\pc$. The implied additional RCGs in the midplane strip increase the total number of RCGs in the $\bb$ by $10\%$ with respect to our fiducial extrapolation. We consider this extrapolation as giving extreme but still reasonable stellar concentration towards the midplane.

The vertical density profiles along the major axis of this variant extrapolation are plotted along with our fiducial ones in \reffigure \ref{fig:alternativeExtrapolation}. In this figure, the different colours indicate the absolute value of the position along the $x'$ axis, from red at the centre to blue at the edges of the 3D map.

\subsubsection{Length scaling}
In the fiducial case we scale our model using the length of the long bar, assuming that this long bar ends at $l = 27\degree$. Even though this assumption seems well founded, such a long bar is in clear tension with previous claims that the pattern speed of the Milky Way bar could be as high as $60\kmskpc$ (see \refsection \ref{section:DiscussionPatternSpeed}). Ongoing work by Wegg et al. (in preparation) shows that the long bar can be reliably traced up to at least $l = 20\degree$. Hence we repeat our experiments by scaling our models on a long bar which would end at $l = 20\degree$ instead of $27\degree$. With the assumed bar angle and distance to the GC the semi-major axis of such a bar would be $3.8\kpc$ long. 

\subsubsection{Snapshot selection}
Our fiducial study is based on initial models which are late evolutionary snapshots of pure disc+halo simulations. Throughout its evolution the bar gives away angular momentum, slows down and builds a strong B/P bulge. As our target density has a very strong peanut shape, late evolutionary stages are a priori more suitable starting points for our modelling. However, when looking at the ratio of the size of the peanut shape to the length of the bar, we found that early evolutionary snapshots better match our target ratio. Consequently we repeat the same modelling analysis using earlier snapshots, taken just after the buckling is complete.

\subsubsection{Bar angle}

The angle between the major axis of the barred bulge and the Sun-GC line has in the past generally been found in the range $20-30\degree$ {\citep{Stanek1997,Bissantz2002,Rattenbury2007a,Nataf2013}}. In this study we assumed an angle of $27\degree$ which is what WG13 measured with an accuracy of $\pm 2\degree$ when making their 3D density map of RCGs in the bulge. Even though this result seems robust we quantify the effect of a different bar angle by experimenting with a bar angle of $32\degree$ ($2.5\sigma$).

\subsubsection{A very robust estimate of the total mass}

\begin{figure}
  \includegraphics{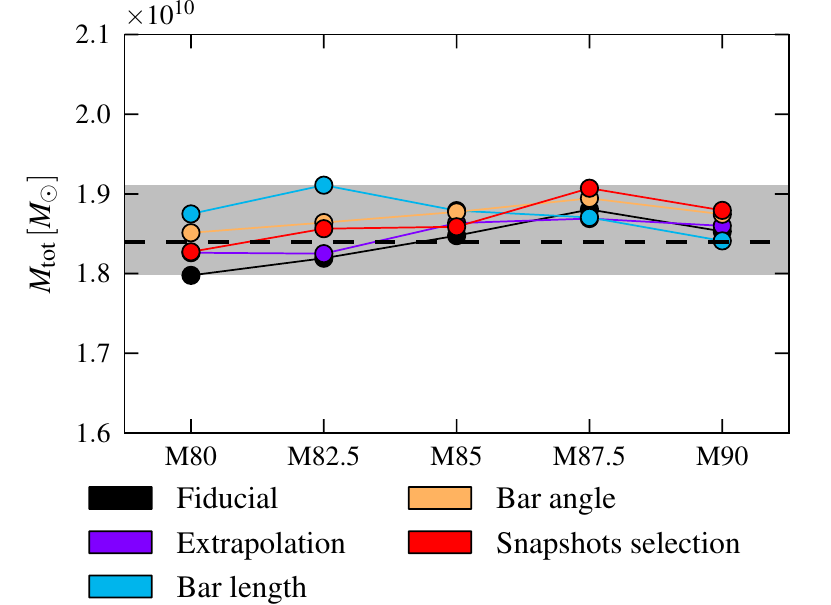}
  \caption{Total mass (stellar + dark matter) of the bulge-in-box for the five models. The black points show our fiducial results already plotted in \reffigure \ref{fig:mass}. The coloured dots show masses obtained when varying some of the model assumptions as stated in the legend.  Only small deviations from our fiducial case are found. The dashed line displays the mean value obtained in the fiducial case while the grey band span the range of all results.}
  \label{fig:systematicsMass}
\end{figure}

\reffigure \ref{fig:systematicsMass} shows the total mass $\Mtot$ as a function of the model for our fiducial case as well as for each variant detailed above. The average total mass in the fiducial case gives an estimate of the total mass of the bulge of $1.84 \times \SM$, shown by the dashed line in \reffigure \ref{fig:mass}. Systematic variations of our four main assumptions have only small effects on the derived value of the total mass as shown in \reffigure \ref{fig:systematicsMass}. As stated in \refsection \ref{section:DynamicalModelsBestF} the uncertainty of the total mass determination due to the discrete sampling of $\F$ is less than $0.01 \times \SM$. The estimated statistical error is $\Delta_{\rm stat}\Mtot= \pm 0.03 \times \SM$. Systematics dominate and are evaluated by simply taking the range of all mass measurements, as showed by the grey band in {\reffigure \ref{fig:mass}}, corresponding to $\Delta_{\rm sys}\Mtot=_{-0.04}^{+0.07}\times \SM \simeq \pm 0.06 \times \SM$. Adding the statistical and systematic error in quadrature, we conclude that the total mass of the bulge-in-box is $1.84 \pm 0.07 \times \SM$. Numerical values of stellar, dark matter and total mass in the $\bb$ for the different models are given in \reftable \ref{table:allmasses}.

\begin{table*}
\caption{Stellar mass $\Ms$, dark matter mass $\MDM$ and total mass $\Mtot$ for all models in the bulge-in-box, in units of $\SM$. The first row refers to our fiducial models while the lower four rows refer to the different systematic variations of the model assumptions detailed in \refsection\ref{section:Systematics}.}
\label{table:allmasses}
  \centering
  \begin{tabular}{l|cc>{\columncolor{light-gray}}c|cc>{\columncolor{light-gray}}c|cc>{\columncolor{light-gray}}c|cc>{\columncolor{light-gray}}c|cc>{\columncolor{light-gray}}c|}
  
    & \multicolumn{3}{c|}{M80}& \multicolumn{3}{c|}{M82.5}& \multicolumn{3}{c|}{M85}& \multicolumn{3}{c|}{M87.5}& \multicolumn{3}{c|}{M90} \\
    & $\Ms$ & $\MDM$ & $\Mtot$ & $\Ms$ & $\MDM$ & $\Mtot$ & $\Ms$ & $\MDM$ & $\Mtot$ & $\Ms$ & $\MDM$ & $\Mtot$ & $\Ms$ & $\MDM$ & $\Mtot$\\
    \hline\hline
Fiducial& 1.27 & 0.53 & 1.80& 1.37 & 0.45 & 1.82& 1.47 & 0.38 & 1.85& 1.59 & 0.29 & 1.88& 1.63 & 0.23 & 1.85\\
Extrapolation& 1.29 & 0.54 & 1.83& 1.36 & 0.47 & 1.83& 1.49 & 0.37 & 1.86& 1.58 & 0.29 & 1.87& 1.63 & 0.23 & 1.86\\
Bar length& 1.32 & 0.55 & 1.87& 1.41 & 0.50 & 1.91& 1.50 & 0.38 & 1.88& 1.57 & 0.30 & 1.87& 1.60 & 0.24 & 1.84\\
Bar angle& 1.30 & 0.55 & 1.85& 1.38 & 0.48 & 1.86& 1.50 & 0.38 & 1.88& 1.60 & 0.29 & 1.89& 1.64 & 0.23 & 1.87\\
Snapshots selection& 1.41 & 0.42 & 1.83& 1.48 & 0.37 & 1.86& 1.56 & 0.30 & 1.86& 1.69 & 0.22 & 1.91& 1.70 & 0.17 & 1.88\\

  \end{tabular} 
\end{table*}

\section{Mass-to-light and mass-to-clump ratios}
\label{section:MtoLAndRCGDensity}
A good estimate of the total mass together with a constraint on the stellar mass would give useful insight into the dark matter mass in the bulge. In this section we constrain the stellar mass in the $\bb$ through the stellar mass-to-light ratio $\Upsilon_K$ and what we call the stellar ``mass-to-clump'' ratio, i.e. the amount of stellar mass per number of Red Clump and Red Giant Branch Bump stars. We first construct predictions from population synthesis models and compare our best mass dynamical models to these predictions. In this way we can relate the stellar Initial Mass Function (IMF) to the dark matter mass of the GB.

\subsection{Population synthesis models}
\label{section:populationsynthesis}

\begin{figure} 
  \includegraphics{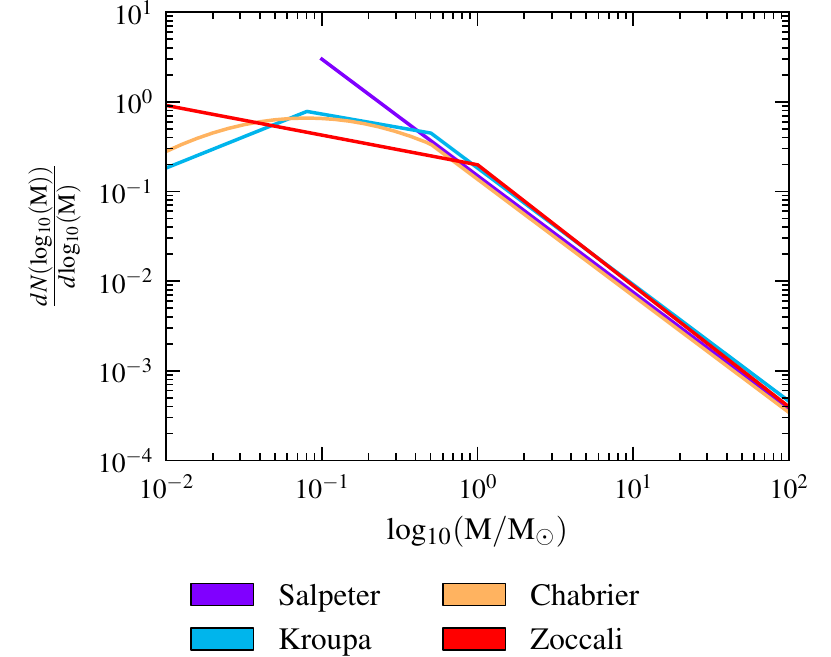}
  \caption{Plot of the logarithmic form of the different IMF used in this study \citep{Salpeter1955,Zoccali2000,Kroupa2001,Chabrier2003}.}
  \label{fig:imf}
\end{figure}

We predict the stellar mass-to-light and mass-to-clump ratios from modelling the evolution of the bulge stellar population. This modelling relies on the three different ingredients.
\begin{enumerate}
 \item An Initial Mass Function (IMF):
 
\noindent We use four IMFs which span the range of reasonable IMFs for the Galactic bulge. Our two extremes are the bottom-heavy Salpeter IMF \citep{Salpeter1955} for masses between $10^{-1}\Msun$ and $10^{2} \Msun$,  and the Zoccali IMF \citep[third entry of their table 3]{Zoccali2000}, the latter being derived specifically from measurements towards the Galactic bulge. In between we use the Kroupa \citep{Kroupa2001} and the Chabrier IMF \citep{Chabrier2003}, which are quite similar. These four IMFs normalized to $1\Msun$ are displayed in \reffigure \ref{fig:imf}. Following \citet{Kroupa2001} we plotted the logarithmic form of the IMF $\xi_{\rm{L}}(\rm{log}_{10}(M)) = M \, \rm{ln}(10)\, \frac{dN}{dM}$, where $\xi_{\rm{L}}(\rm{log}_{10}(M))\, d\rm{log_{10}} (M)$ corresponds to the fraction of stars with mass between $\rm{log_{10} (M)}$ and $\rm{log_{10} (M)} + d\,\rm{log_{10} (M)}$ and $\rm{M}$ is expressed in $\rm{M}_{\odot}$.
 
 \item \label{itm:isochrone} A set of isochrones:
 
\noindent We choose the solar metallicity and $\alpha$-enhanced \textsc{BaSTI} isochrones \citep{Pietrinferni2004} and assume a single age population of $10\, \rm{Gyr}$. As shown later the choice of the age has a small effect and is therefore not critical.
 
 \item \label{itm:remnants} A way to treat stellar remnants:
 
\noindent Stars that evolve beyond their isochrones have to be turned properly into white dwarfs, neutron stars or black holes. We use the choices described in \citet{Maraston1998}.
\end{enumerate}

Using these three ingredients, we first construct a luminosity function $\Phi(M_K)$ in units of $\rm{mag^{-1}\Msun^{-1}}$ by evolving $1\Msun$ through the set of isochrones for a given age, according to the considered IMF mass distribution. After renormalization of the luminosity function to a remaining mass of $1\Msun$ (evolved stellar population + stellar remnants), the synthesized stellar mass-to-light ratio in the K-band $\Upsilon_K$ is given by

\begin{equation}
 \Upsilon_K = \left( \int \Phi(M_K) 10^{-0.4 (M_K-M_{K_{\odot}})} \, dM_K \right)^{-1}.
\end{equation}
We checked our mass-to-light ratio predictions by reproducing the work of \citet{Maraston1998} and \citet{Percival2009} and found a very good agreement with both of them for the old population considered here.

In order to compute the stellar mass-to-clump ratio we use the same technique as in WG13. We fit an exponential background plus two Gaussians to the previously derived luminosity function. The two Gaussians represent the RCGs and the Red Giant Branch Bump (RGBB), as in WG13. This mass-to-clump ratio includes RGBB stars as well for two reasons. First because for some ages, the RCGs and RGBB have the same luminosity and are therefore indistinguishable. Secondly because WG13 made their map by fitting these two Gaussians under the assumption that the RGBB represents $20\%$ of the Red Clump. Hence the total number of RCGs and RGBB stars is a priori better constrained than the number of RCGs only. As the luminosity function was renormalized to a remaining mass of $1 \Msun$, the number of stars contained in the two fitted Gaussians directly leads to the mass-to-clump ratio, denoted as $M/\rm{n}_{\rm{RC+RGBB}}$.

\subsection{Mass-to-light ratio}
\label{section:MtoL}

The COBE/DIRBE instrument provides us with $K$-band measurements in many bulges fields. Here we use the data from \citet{Drimmel2001} who removed point sources by applying a median filter to the original ``Zodi-Subtracted Mission Average'' map \citep{Kelsall1998}. In order to correct for Galactic extinction, we use the extinction map presented in WG13, also derived from $K$-band data. This map covers the inner $l \in [-10\degree,5\degree], b \in [-10\degree,10\degree]$ of the bulge and has a resolution of $1'$ which is much finer than the $42' \times 42'$ COBE/DIRBE field size. We correct the COBE/DIRBE data by using the mean extinction on each COBE/DIRBE pointing. We ignore the fields in the inner $|b|< 2 \degree$ because the extinction is too high to make a reliable correction. We apply a disc contamination correction to the remaining data points as follows. The average disc contamination as a function of latitude is evaluated using the surface brightness profile in two $0.5\degree$ wide strips along the $b$ direction, located at $l = \pm15\degree$. At these large longitudes the Galactic bulge is no longer important and the flat long bar does not contribute significantly to the surface brightness for $|b|\geq 2\degree$. Hence the average surface brightness profile of the strips is mostly due to the disc component. By assuming that this disc vertical surface brightness profile stays constant at all longitudes inside $|l|<15\degree$ we can estimate and then remove the disc contamination from the COBE data. Finally this provides us with about $2800$ extinction and foreground corrected surface brightness measurements towards the Galactic bulge which can be compared to our models.

To compute the surface brightness of our particle models we convert scaled stellar particle weights $w_i$ into $K$-band luminosity $L_i = w_i \Upsilon_K^{-1}$, assuming a constant but still unknown stellar mass to-light ratio $\Upsilon_K$ in the bulge. $\Upsilon_K$ will be determined by matching our model surface brightness to the COBE/DIRBE $K$-band data.
The extinction-free model surface brightness in some field $\Sigma_j (l,b)$ as one would see from the Sun's location is given by
\begin{equation}
 \Sigma_j(l,b) = \frac{1}{\Delta \Omega_j} \sum_i \frac{w_i . \Upsilon_K^{-1}}{r_i^2} \delta_j(r_i)
 \label{equation:modelLight}
\end{equation}
where $\Delta \Omega_j $ is the solid angle of the considered field, $r_i$ is the distance of the particle $i$ from the Sun in $\kpc$, and $\delta_j(r_i)$ is a suitable selection function that we describe below in \refequation \ref{equation:selectionMtoL}. Our models have been matched to the MW inside the $\bb$ but no attempts have been made to match the model discs to the MW disc. In order to reduce the uncertainty due to the disc contribution we evaluate for each COBE field the fraction of model light which comes from particles located in the $\bb$. We remove from the analysis all fields where this fraction is lower than $90 \%$, i.e. all fields where not directly constrained particles contribute more than $10 \%$ of the model surface brightness. Given that a single particle can theoretically dominate the surface brightness by being arbitrarily close to the Sun's location, we also remove nearby disc particles from the analysis in the remaining fields by taking the following selection function:

\begin{equation}
  \label{equation:selectionMtoL}
 \delta_j (r_i) = 
 \begin{cases}
   1 & \text{if } i \in \text{field }j \text{ and } r_i \geq 3 \kpc \\
   0 & \text{otherwise}
  \end{cases}
\end{equation}
We then apply the same foreground contamination correction to the model surface brightness as for the data, using the two strips at  $l = \pm15\degree$. We checked that our results do not depend on the exact form of the selection function, indicating that the foreground contamination has been properly removed.

Finally, we compute the mass-to-light ratio as stated in \refequation \ref{equation:modelLight} independently for all remaining COBE fields and average the results. The statistical error in the mean mass-to-light ratio is very low due to the large number of COBE fields so that systematics dominate. We evaluate systematic effects by repeating this analysis for the four variants described in \refsection \ref{section:Systematics} of the considered model. The full range of values is then taken as the systematic error.

\begin{figure}
  \includegraphics{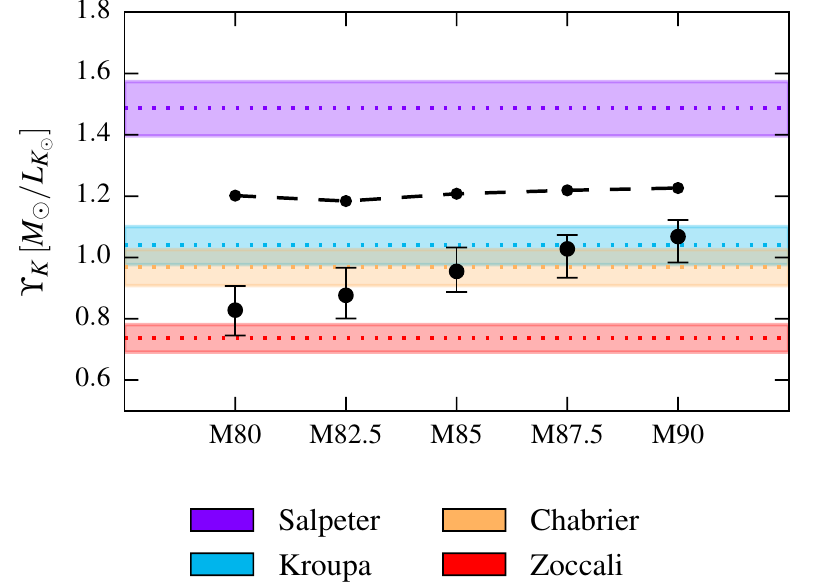}
  \caption{Stellar mass-to-light ratio in the $K$ band for our five best dynamical models of the bulge in different dark matter haloes. The model errors shown are dominated by systematic effects. The different coloured lines indicate predictions from different IMFs as stated in the legend. The black dashed line is an estimate of the highest allowed mass-to-light ratio obtained by turning all dark matter of the $\bb$ into luminous matter. This allows us to rule out the Salpeter IMF for the Galactic bulge with age $10 \Gyr$.}
  \label{fig:MtoL}
\end{figure}

\reffigure \ref{fig:MtoL} shows the mean value of $\Upsilon_K$ and its associated systematic error for the five best dynamical models with different dark matter haloes. These mass-to-light ratios are compared to the predictions from different IMFs shown as the coloured lines, for a single $10\Gyr$ age population. The coloured strips span the range of values for ages between $9$ and $11\Gyr$. With the total mass and integrated light fixed, the models with low dark matter mass (M90 for example) have higher stellar mass-to-light ratios. We see that most of our models are in agreement with predictions from a Kroupa or Chabrier IMF which are somewhat similar, the Kroupa IMF predicting slightly higher mass-to-light ratios than the Chabrier IMF. Only the most dark matter dominated model M80 approximately matches the predictions from the Zoccali IMF, which is a priori the best candidate IMF because it was measured directly from the stellar luminosity function of the bulge. In order to agree with the Zoccali IMF about $40\%$ dark matter mass is required in the $\bb$.

In addition, the dynamical models rule out a Salpeter IMF for a bulge population with age $10\Gyr$. The dashed line in \reffigure \ref{fig:MtoL} represent the highest possible stellar mass-to-light ratio which would be obtained if all the dark matter in the bulge was turned into stars with the same density distribution as the stellar component of the bulge in our fiducial models. For all our models, the mass-to-light predictions for a Salpeter IMF for a $10\Gyr$ Galactic Bulge are higher than this extreme mass-to-light ratio by at least three times the model error, showing that it is too bottom-heavy.

\subsection{Mass-to-clump ratio}
\label{section:RCGDensity}

\begin{figure} 
  \includegraphics{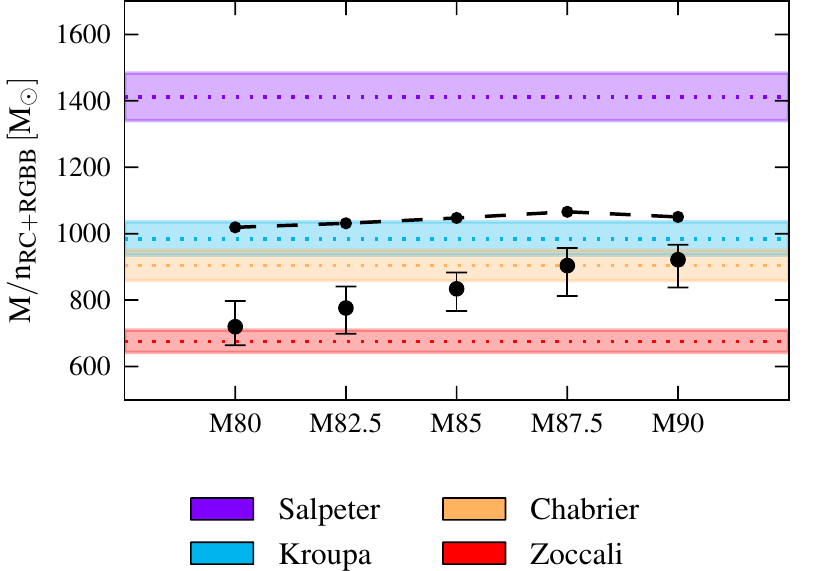}
  \caption{Mass-to-clump ratio, $M/\rm{n}_{\rm{RC+RGBB}}$ for our five best dynamical models of the bulge in different dark matter haloes. The coloured lines indicate predictions from different IMFs as stated in the legend. The black dashed line is an estimate of the highest allowed mass-to-clump ratio obtained by turning all dark matter of the $\bb$ into luminous matter.}
  \label{fig:nRC}
\end{figure}

The mass-to-clump ratio is also a useful quantity to relate stellar mass and stellar population. With RCGs being approximate standard candles with standard colours, issues like foreground contamination and dust extinction are easier to solve when computing a mass-to-clump ratio than a mass-to-light ratio. The number of RCGs in the $\bb$ was computed directly from our fiducial extrapolation of the density map of WG13. Using the variant extrapolation presented in \refsection \ref{section:Systematics} would increase this number by only $10\%$. In order to include the RGBB in this calculation, we add the $20\%$ fraction assumed in the derivation of the 3D map by WG13. The computation of $M/\rm{n}_{\rm{RC+RGBB}}$ is then straightforward from the stellar mass determination of \refsection \ref{section:MassEvaluation}. Results are shown in \reffigure \ref{fig:nRC} for all dark matter models. The errors plotted in \reffigure \ref{fig:nRC} are systematic, determined from the four variant assumptions described in \refsection \ref{section:Systematics}. Again the different colour strips indicate predictions of different IMFs for ages between $9$ and $11\Gyr$.

We reach similar conclusions from the mass-to-clump ratio as from the mass-to-light ratio. This indicates that the issues of extinction and foreground contamination present in the computation of the mass-to-light ratio have been treated correctly. A $35\%$ dark matter fraction in the $\bb$ is needed to match predictions from the Zoccali IMF and low dark matter models would need a Chabrier IMF to match the predictions for a $10\Gyr$ old population. Once again the Salpeter IMF over-predicts the mass-to-clump ratio by about $50\%$ and can therefore be ruled out for a $10\Gyr$ GB.

\section{Peanut and X-shape structures of the Galactic bulge}
\label{section:Xshape}

In this section we study the structural properties of the GB further, based on the 3D density map of RCGs from WG13. We first illustrate and discuss its X-shape in a similar way as usually done for external galaxies. We then use a 3D photometric diagnostic to quantify the fraction of stellar mass involved in the peanut shape of the GB.

\subsection{The photometric X-shape}
\label{section:XshapeMedianFilter}

\begin{figure}
  \includegraphics{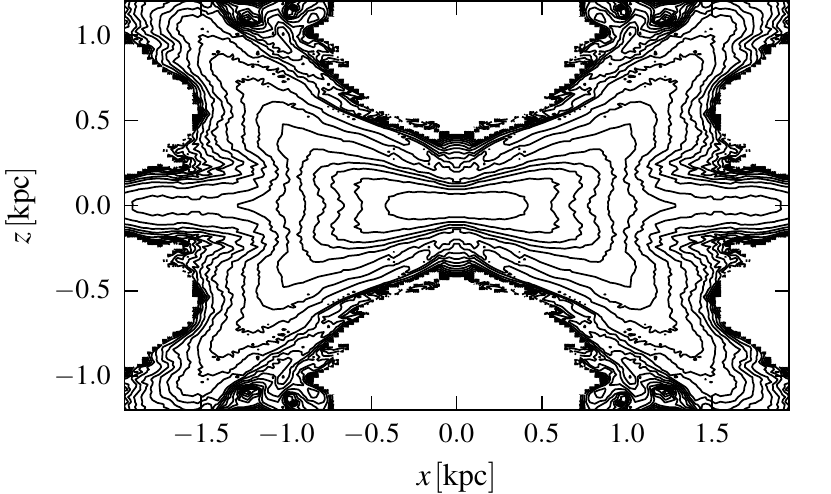}
  \caption{X-shape structure of the Galactic Bulge obtained by unsharp masking as described in \citet{Bureau2006}. The Galactic bulge has a so-called off-centred X-shape.}
  \label{fig:medianFilter}
\end{figure}

Observations of external galaxies have revealed that Box/Peanut bulges exist with a variety of shapes. In order to highlight the internal structure of external edge-on B/P bulges, a common practice is to use unsharp masking techniques as described by \citet{Bureau2006}. They applied a median filter to images of $30$ edge-on spirals and removed the smoothed images from the original ones. This reveals what is called the X-shape. \citet{Bureau2006} proposed a classification of external B/P bulges based on the properties of this X-shape. Where would the Galactic bulge appear in such a classification? To answer this question, we apply a median filter with kernel size $500\pc$ to the side-on projection of the 3D density of RCGs from WG13 and remove it from the original projection. The positive residuals revealing the X-shape of the GB are shown in \reffigure \ref{fig:medianFilter} with contours spaced by a third of a magnitude.

\reffigure \ref{fig:medianFilter} shows that the Galactic bulge has an off-centred X-shape structure, its two arms crossing the major axis about $500 \pc$ away from the centre. This feature is identified in $50\%$ of external edge-on B/P bulges \citep{Bureau2006}.

\subsection{The mass of the peanut shape}
\label{section:XshapePhotometricDiagnostic}
Consistent with the strong X-shape, the Milky Way bulge in the side-on map from WG13 shows a very prominent peanut shape. An interesting issue is the amount of stellar mass involved in this feature. This was already addressed by \citet{Li2012} who applied a technique somewhat similar to unsharp masking to the side-on projection of the model of \citet{Shen2010}. They fitted ellipses to the isophotes of the side-on projection, modelled the light projected by these elliptical isophotes and removed the modelled light from the original image. Doing so revealed a centred X-structure in the model accounting for about $7\%$ of the bulge stellar mass. Because their model had been shown to give a good representation of the \brava kinematic data, they then concluded that the stellar mass involved in the peanut shape of the GB was similarly about $7\%$.

Since we now have a direct measurement of the 3D density of RCGs in the bulge, we can perform a similar photometric analysis directly on the 3D density map. Consider the density profile $\rho(0,0,z)$ along the minor axis of the bar. For each particular value of $z$, we evaluate the 3D isodensity surface with density $\rho(0,0,z)$. Then we look for the most voluminous ellipsoid we can find which fits \emph{inside} this particular 3D isodensity surface. This search is done under the constraint that the ellipsoid is centred on the centre of the bulge, the principal axes are aligned with the principal axes of the bulge, the semi-principal length along the vertical axis is fixed to $|z|$, and the ellipsoid stays inside the isodensity surface considered. By doing so for all $z$ we construct a family of ellipsoidal isodensity surfaces. We compute the 3D density arising from this ellipsoidal isodensity family and remove it from the original 3D map. The residuals correspond to the ``non-ellipsoidal'' part of the bulge density. As all our ellipsoids are truly inside their corresponding isodensity surfaces of the original map, we are assured that the residual map is positive at all points.

To reduce the effect of measurement errors in the original map of WG13, we actually do this analysis using model M80, which through \nmagic fitting of the density gives a smoother density map which is everywhere within the errors of the original map. We note that the results do not depend on which initial model is used for the density fit.

\begin{figure}
  \includegraphics{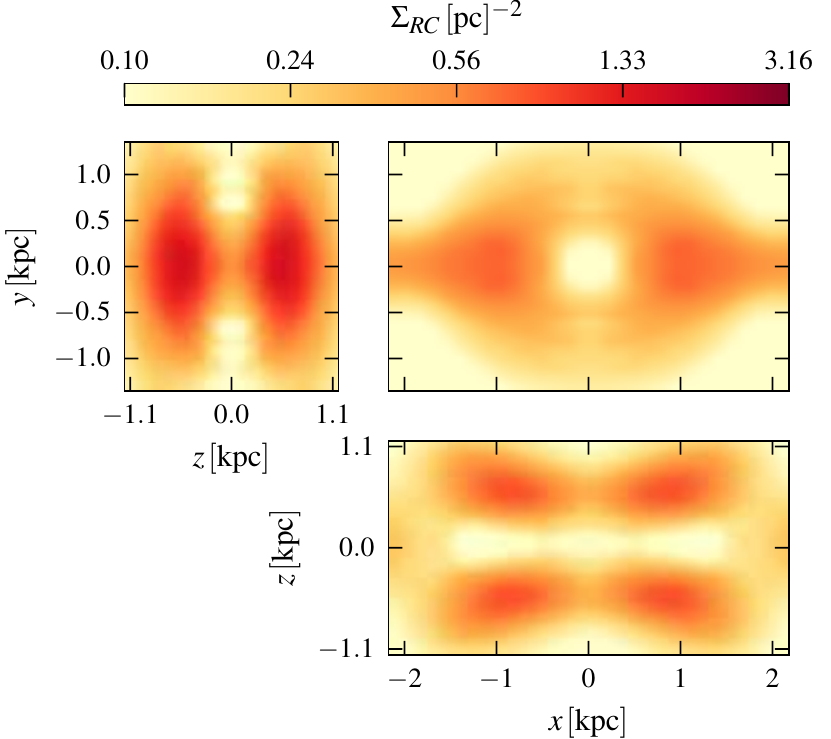}
  \caption{``Non-ellipsoidal'' residual density of RCGs in the density map of \citet{Wegg2013}. The residual density accounts for $24\%$ of the RCGs in the bulge. Plotting conventions are the same as in \reffigure \ref{fig:RCGDensity}.}
  \label{fig:XShape}
\end{figure}

The residual map is plotted in projection in \reffigure \ref{fig:XShape}. The four lobes responsible for the peanut shape of the bulge are clearly visible in the side-on view for $|z|>300\pc$. For $|z|<300\pc$ the peanut is not prominent enough with respect to its surroundings to make the density shape deviate from ellipsoidal shape. The stellar mass involved in this residual peanut shape is about $24\%$ of the total stellar mass of the bulge. This figure probably underestimates the real amount of mass involved in the peanut structure as one expects the orbits responsible for this structure to also visit the strip inside $|z|<300\pc$ and therefore to contribute to the ellipsoidal density that was removed here.

We can estimate an error on the $24\%$ by applying the same procedure to each of the modelled five variant maps of WG13. We find that the non-ellipsoidal residuals account for $24_{-4}^{+5} \%$ of the stellar mass of the bulge. By applying our diagnostic in a two dimensional version to the side-on projection of the density, we find that the 2D residuals sum to only $11_{-1}^{+1} \%$ of the stellar mass of the bulge, showing that 2D determinations tend to underestimate the mass in the peanut structure.

Our peanut mass fraction of $24_{-4}^{+5} \%$ results from an observational definition of the peanut shape: a deviation from ellipsoidal density shape. A more physical definition of the peanut structure would be to identify the different orbits responsible for its shape in side-on projection. We are currently working on such an orbit-based characterization of the peanut shape which will be presented in a companion paper.

\section{Discussion}
\label{section:Discussion}
\subsection{Dynamical mass of the bulge}
The Galactic B/P bulge transits into a longer two-dimensional bar; therefore in this paper we use a simple definition for determining the dynamical mass of the bulge. We use the three-dimensional box $\pm 2.2 \times \pm 1.4 \times \pm 1.2\kpc$ as our bulge volume in which the RCGs density was determined by \citet{Wegg2013} and which contains most of the three-dimensional bulge part of the bar. The total mass of this bulge-in-box ($\bb$) is accurately determined by the dynamical models, ${\Mtot} = 1.84 \pm 0.07 \times \SM$. This value is essentially independent of the dark matter mass fraction in the bulge throughout our models, and the estimated error includes both statistical and systematic uncertainties ; the systematic part is determined from varying the modelling assumptions around our fiducial model.

Dynamical models have previously been used to estimate the mass of the bulge. Using the Schwarzschild method, \citet{Zhao1994} built a self-consistent model of the bar/bulge and found a total bulge mass of $2 \times \SM$ while \citet{Kent1992} found a mass of $1.8 \times \SM$ by modelling an oblate isotropic rotator with a constant mass-to-light ratio. By studying gas dynamics in the potential of the model from \citet{Bissantz2002} obtained by deprojecting the COBE luminosity distribution, \citet{Bissantz2003} determined the circular velocity at $2.2 \kpc$ to be $190\kms$. Once converted to mass under the assumption of spherical symmetry, this leads to a total bulge mass of about $1.85 \times \SM$. All these results compare well with our estimate of the mass of $1.84 \pm 0.07 \times \SM$, especially when considering the difficulty of a precise definition of the bulge.

An independent way to obtain a dynamical mass is from the virial theorem. Such studies lead to quite different values depending on the assumed pattern speed and bar angle, from $1.6 \times \SM$ \citep{Han1995}, up to $2.8 \times \SM$ \citep{Blum1995} obtained for a high pattern speed ($81\kmskpc$). The use of the virial theorem has two weaknesses. First, it relies on an estimation of the total kinetic energy, which cannot be measured accurately from line-of-sight data only. Secondly, it assumes that the bulge on its own is a system in equilibrium, whereas it is in fact part of a bar embedded in a disc.

Traditionally, the contribution of dark matter to the mass in the bulge region has been considered unimportant. In this case, a bulge mass can be estimated simply from the stars. For example,
\citet{Dwek1995} estimated the stellar mass of the bulge at $1.3 \times \SM$ from the COBE luminosity and a Salpeter IMF.  However, neglecting dark matter in the bulge is not a good approximation as we have seen in \refsection \ref{section:mass}.

\subsection{Stellar and dark matter mass in the bulge}
\label{section:DiscussionDarkHalo}

Because the total mass of the bulge-in-box ($\bb$) is well determined, knowing the stellar mass independently would give insight into the amount of dark matter in the central part of the Galaxy.  As shown in \refsection \ref{section:MtoLAndRCGDensity}, the stellar mass can be constrained by comparing mass-to-light and mass-to-clump ratio measurements with stellar population synthesis predictions.  However, for this the choice of the IMF is crucial: candidate IMFs disagree within a factor of two in their prediction of the stellar mass-to-light ratio, so a reliable measurement of the IMF in the GB is needed.  Currently, the Zoccali IMF \citep{Zoccali2000} measured from the bulge star luminosity function in a field located near $(l,b) = (0\degree, -6\degree)$ is the favoured IMF for the GB. For this case, the predicted mass-to-light ratio from isochrones (\refsection \ref{section:populationsynthesis}) gives a stellar mass for the $\bb$ of $1.12 \times \SM$. From our modelling, both the mass-to-light and the mass-to-clump ratio independently agree that models with fairly high dark matter mass fraction are required to provide the remaining part of the dynamical mass in the $\bb$.  These models predict a dark matter mass in the $\bb$ of $M_{DM} \sim 0.7 \times \SM$, accounting for about $40\%$ of the total mass of the $\bb$.

Further insight on the dark matter part of the $\bb$ mass can be obtained from proper motions. In particular, the proper motion dispersions in the $b$ direction directly constrain the derivative of the potential along the vertical direction, and therefore the total mass concentration towards the plane. We showed in \refsection \ref{section:PM} that our proper motion predictions $\sigma_b$ are $10\%$ to $20\%$ lower than the data. Increasing the stellar mass concentration towards the plane as in \refsection \ref{section:variantExtrapolation} can indeed increase $\sigma_b$, but only at percent level which is not significant enough. It is also possible that a systematic effect is present in the data, e.g., due to extinction or incompleteness of the sample. Before we can use the proper motion data to measure the mass concentration and dark matter content of the GB, these possible systematic effects in the data need to be better understood.  A more detailed study of the proper motion constraints is part of our ongoing work to make a more complete model of the Galactic bulge and long bar.

\subsection{Pattern speed of the MW bar/bulge}
\label{section:DiscussionPatternSpeed}

Our dynamical models, based on the RCGs density from \citet{Wegg2013} together with the \brava kinematic data for the bulge stars, imply a pattern speed of the MW bar-bulge of $25-30\kmskpc$ (see \refsection \ref{section:DynamicalModelsPatternSpeed}). This result remains unchanged when varying the modelling assumptions, for example when we consider a shorter bar, as detailed in \refsection \ref{section:Systematics}. Comparing with the composite rotation curve from \citet{Sofue2009}, this would place the corotation value of the bar just inside the solar circle, and give a ratio of corotation over bar half length between $1.5$ and $1.8$.  The MW would then belong to the so-called slow rotators. Slow rotators are quite rare for external galaxies in general \citep{Aguerri1998}, but fairly common among SBc galaxies \citep{Rautiainen2008}.

There have been a quite a number of pattern speed measurements in the MW by other techniques. The bulk of the measurements indicate a fast bar, but some studies suggest a lower pattern speed (see \cite{Gerhard2011} for a review). Direct measurement by \citet{Debattista2002} using a variant of the Tremaine \& Weinberg method \citep{Tremaine1984} leads to the high value of $\Omega_{\rm p} = 59 \pm 5 \kmskpc$ but depends strongly on the radial velocity of the local standard of rest towards the GC.

Numerous indirect measurements of the bar pattern speed have been obtained by matching some of the observed features in the position-velocity diagrams for H{\scriptsize I} and CO to gas-dynamical model predictions. Most of these studies argue for a high pattern speed, such as \citet{Englmaier1999} who found $\sim 60 \kmskpc$, \citet{Fux1999} who obtained $\sim 50 \kmskpc$, and \citet{Bissantz2003} with $55 - 65 \kmskpc$. However, \citet{Weiner1999} and \citet{Rodriguez-Fernandez2008} obtained lower values of respectively $42 \kmskpc$ and $30 - 40 \kmskpc$.

High pattern speeds were also obtained by explaining the stellar kinematics of nearby disc stars with the dynamical effects of the Outer Lindblad Resonance (OLR) of the bar. By interpreting the
bimodality of the velocity distribution in the solar neighbourhood in this way, \citet{Dehnen2000} found $\Omega_{\rm p} = 53 \pm 3 \kmskpc$. \citet{Antoja2014} presented an analytical model of the effect of the OLR in the velocity distributions of stars at different radii and showed that they could reproduce measurements of the Hercules stream for a bar pattern speed of $\Omega_{\rm p} = 54 \pm 0.5 \kmskpc$.

The low pattern speed value found from the RCGs and \brava data is consistent with \citet{Rodriguez-Fernandez2008} and \citet{Long2013} but not with the majority of these measurements.
We note that models based on variants of the RCGs density map as well as models constrained by a model B/P bulge density independent of the RCGs measurement \citep{Martinez-Valpuesta2012} give similarly low values. The \brava kinematics are in agreement with the \argos kinematics \citep{Ness2013a} despite the quite different selection functions of both surveys. In our modelling, the details of the selection function assumed for the \brava data were not important. Therefore our value measured from modelling the bulge kinematics appears robust. 

The highest pattern speeds reported in the literature are also in conflict with other data. Using the composite rotation curve from \citet{Sofue2009}, a pattern speed of $\Omega_{\rm p} = 55 \kmskpc$ would place corotation at $3.7 \kpc$. Because the long bar cannot extend beyond corotation \citep{Contopoulos1980}, this is incompatible with the apparent end of the long bar at $l = 27 \degree$ \citep{Hammersley2000, Cabrera-Lavers2008}. Even if we relax the hypothesis that the long bar ends at $l = 27\degree$, recent star counts from the \ukidss survey (Wegg et al.\ in preparation) show that it can be reliably traceable at least out to $l = 20\degree$. This gives a lower bound of the half-length of the bar of $3.8\kpc$, still in tension with a pattern speed of $55 \kmskpc$.

We conclude that despite of the many effort made by different groups, the question of the pattern speed of the Galactic bar remains an unsolved issue. One way to settle this question by dynamical modelling would be to include data that more accurately constrain the long bar component, which is one of our future goals.

\section{Conclusion}
\label{section:Conclusion}

In this work we have presented a set of self-consistent dynamical models of the Galactic bulge with different dark matter haloes, which match recent data on the spatial distribution and kinematics of bulge stars. We started with a family of N-body models of barred discs with B/P bulges, evolved from near-equilibrium stellar discs embedded in live dark matter haloes.
We then fitted these models to the recent 3D density measurements of Red Clump Giants in the bulge from \citet{Wegg2013}, as well as to the \brava kinematic data of \citet{Kunder2012} in multiple bulge fields, using the \nmagic Made-to-Measure method.

From this modelling, we obtain an accurate and robust estimate of the total mass (stellar and dark matter) of the bulge in the RCGs box, of ${\Mtot} = 1.84 \pm 0.07 \times \SM$. We also find a low pattern speed of about $25-30 \kmskpc$, which with the measured rotation curve places the Milky Way among the slow rotators ($\R \geq 1.5$). We compute the mass-to-light and mass-to clump ratios and compare them with theoretical predictions from population synthesis models using different IMFs. We show that the Salpeter IMF \citep{Salpeter1955} is ruled out for a $10 \Gyr$ old bulge population. We find that a relatively high dark matter mass fraction in the bulge is needed in order to match predictions from the IMF inferred from the stellar luminosity function in the upper bulge \citep{Zoccali2000}, $\sim 40\%$ or $\MDM \sim 0.7 \times \SM$. 
In addition we study the X-shape of the Galactic bulge and find an off-centred X-shape, common in external B/P bulges \citep{Bureau2006}. By a three-dimensional analysis of the isodensity surfaces of the RCGs density we find that the peanut-shaped deviations from ellipsoidal shape account for $24_{-4}^{+5} \%$ of the bulge stellar mass, significantly larger than the previous estimate of $7\%$ of \citet{Li2012}.

\section*{Acknowledgements}
We gratefully acknowledge the work of Flavio De Lorenzi in writing the initial version of \nmagic, Mat\'ias Bla\~na for his help and advice in setting up our initial N-body models, and Jerry Sellwood and Monica Valluri for making their potential solver available. We thank the anonymous referee for a careful reading of the paper.

\bibliographystyle{mn2e}

\bsp
\label{lastpage}
\end{document}